\documentclass[structabstract]{aa}

\usepackage{graphicx}
\usepackage{subfigure}
\usepackage{float}
\usepackage{psfrag}
\usepackage{amsmath}
\usepackage{amssymb}
\usepackage{ulem}
\usepackage{natbib} 
\usepackage{txfonts}

\begin{document}
   \title{Multi-frequency imaging of the galaxy cluster Abell 2163 using the
   Sunyaev-Zel'dovich Effect} 

   \subtitle{}

   \author{
     M. Nord\inst{1 \and2}\fnmsep\thanks{\email{mnord@astro.uni-bonn.de}} \and 
     K. Basu\inst{1 \and2} \and
     F. Pacaud\inst{1} \and 
     P. A. R. Ade\inst{3} \and
     A. N. Bender\inst{4} \and 
     B. A. Benson\inst{5} \and
     F. Bertoldi\inst{1} \and
     H.-M. Cho\inst{6} \and 
     G. Chon\inst{1 \and2} \and
     J. Clarke\inst{5} \and 
     M. Dobbs\inst{7} \and 
     D. Ferrusca\inst{5} \and 
     N. W. Halverson\inst{4} \and 
     W. L. Holzapfel\inst{5} \and
     C. Horellou\inst{8} \and
     D. Johansson\inst{8} \and
     J. Kennedy\inst{7} \and 
     Z. Kermish\inst{5} \and 
     R. Kneissl\inst{2} \and
     T. Lanting\inst{7 \and9} \and 
     A. T. Lee\inst{5 \and10} \and 
     M. Lueker\inst{5} \and 
     J. Mehl\inst{5} \and 
     K. M. Menten\inst{2} \and 
     T. Plagge\inst{5} \and 
     C. L. Reichardt\inst{5} \and 
     P. L. Richards\inst{5} \and 
     R. Schaaf\inst{1} \and 
     D. Schwan\inst{5} \and 
     H. Spieler\inst{10} \and 
     C. Tucker\inst{3} \and
     A. Weiss\inst{2} \and 
     O. Zahn\inst{5}
   }

   \institute{
     Argelander Institute for Astronomy, Bonn University, Bonn, Germany
     \and
     Max Planck Institute for Radioastronomy, 53121 Bonn, Germany
     \and
     School of Physics and Astronomy, Cardiff University, CF24 3YB Wales, UK
     \and
     Center for Astrophysics and Space Astronomy, University of Colorado,
     Boulder, CO, 80309, USA 
     \and
     Department of Physics, University of California, Berkeley, CA, 94720, USA
     \and
     National Institute of Standards and Technology, Boulder, CO, 80305, USA
     \and
     Physics Department, McGill University, Montreal, Canada H2T 2Y8
     \and
     Onsala Space Observatory, Chalmers University of Technology, 43992
     Onsala, Sweden
     \and
     D-Wave Systems Inc., Burnaby, Canada V5C 6G9
     \and
     Lawrence Berkeley National Laboratory, Berkeley, CA, 94720, USA
   }

   \date{Received ... ; accepted ...}

  \abstract
  {Observations of the Sunyaev-Zel'dovich effect (SZE) from galaxy clusters
    are emerging as a powerful tool in cosmology. Besides large cluster
    surveys, resolved SZE images of individual clusters can shed light on the
    physics of the intra-cluster medium (ICM) and allow accurate
    measurements of the cluster gas and total masses.}
  {We used the APEX-SZ and LABOCA bolometer cameras on the APEX telescope
    to map both the decrement of the SZE at 150 GHz and the increment at 345
    GHz toward the rich and X-ray luminous galaxy cluster Abell 2163 at
    redshift 0.203. The SZE images were used, in conjunction with archival
    XMM-Newton X-ray data, to model the radial density and temperature
    distribution of the ICM, as well as to derive the gas mass fraction in the
    cluster under the assumption of hydrostatic equilibrium.}
  {We describe the data analysis techniques developed to extract the faint and
    extended SZE signal.  We used the isothermal $\beta$ model to fit the SZE
    decrement/increment radial profiles.  We performed a simple, non-parametric
    de-projection of the radial density and temperature profiles, in
    conjunction with X-ray data, under the simplifying assumption of spherical
    symmetry.  
    We combined the
    peak SZE signals derived in this paper with published SZE measurements of
    this cluster to derive the cluster line-of-sight bulk velocity and the
    central Comptonization, using priors on the ICM temperature.}
  {We find that the best-fit isothermal model to the SZE data is consistent
    with the ICM properties implied by the X-ray data, particularly inside the
    central 1 Mpc radius.  Inside a radius of $\sim$1500 kpc from the cluster
    center, the mean gas temperature derived from our SZE/X-ray joint analysis
    is
    {$10.4\pm1.4$} keV. {The error budget for the derived
      temperature profile is dominated by statistical errors in the 150 GHz
      SZE image.}  From the isothermal analysis combined with previously
    published data, we find a line-of-sight peculiar velocity
    consistent with zero;
    {$v_r=-140\pm460$} km/s, and a central Comptonization
    {$y_0=3.42\pm0.32 \times 10^{-4}$} for
    Abell 2163.}
  {Although the assumptions of hydrostatic equilibrium and spherical symmetry
    may not be optimal for this complex system, the results obtained under
    these assumptions are consistent with X-ray and weak-lensing measurements.
    This shows the applicability of the simple joint SZE and X-ray
    de-projection technique described in this paper for clusters with a wide
    range of dynamical states. }


\maketitle


\section{Introduction}

The Sunyaev-Zel'dovich effect (SZE, Sunyaev \& Zel'dovich 1970, Birkinshaw
1999) provides a powerful probe of the large-scale structure in the Universe
by imprinting the thermal energy of clusters of galaxies on the cosmic
microwave background (CMB). The inverse Compton scattering of CMB photons by
the hot intra-cluster medium (ICM) has a characteristic frequency dependence
and a power spectrum that is very different from that of the primary CMB
anisotropies. 

A number of blind SZE imaging surveys are in progress (e.g. Ruhl et al. 2004,
Kosowsky 2006) with the aim of tracing the large-scale structure of the
Universe through the detection of galaxy clusters, and the first blind SZE
detections have recently been reported (Staniszewski et al. 2009).

With sufficiently high resolution, the SZE can be used to map the pressure
structure of individual clusters. Once combined with information on the X-ray
surface brightness, the SZE thus provides constraints on gas temperatures and
total mass distributions inside clusters. Such constraints are free of the
potential biases of X-ray spectroscopy.  A further comparison with
weak-lensing maps yields insight into the dynamical state of the cluster by
checking the validity of the hydrostatic equilibrium condition.  Joint
de-projection methods using X-ray and SZE maps have been proposed (e.g. Lee \&
Suto 2004, Puchwein \& Bartelmann 2006), but so far their implementation using
real SZE data has been limited to unresolved or barely resolved SZE maps (e.g.
De Filippis et al. 2005). {Kitayama et al. (2004) presented the first
  tempertaure deprojection of a cluster using a combined SZE and X-ray
  analysis. However, this analysis was limited to using a parametric model for
  the SZE signal.} One main objective of this paper is to show the potential
of such a de-projection analysis using high signal-to-noise resolved SZE maps
{without the limitations of parametric models}.

Abell 2163 is a hot, X-ray luminous galaxy cluster at $z=0.203$, with a mean
X-ray temperature in the central region of $T_{\mathrm{X}}=12^{+1.3}_{-1.1}$
keV (Markevitch \& Vikhlinin 2001). It has a relatively large angular extent,
with a virial radius estimated from weak-lensing mass modeling to be on the
order of 15 arcminutes (Radovich et al. 2008), making the primary CMB
anisotropies a major source of systematic uncertainty in the SZE measurements.
Detailed X-ray observations suggest that Abell 2163 is a merger system (Elbaz
et al. 1995, Govoni et al. 2004), with asymmetric X-ray temperature structures
and strong radio halos. The merger scenario is also supported by optical
observations, most notably from the presence of two bright cD galaxies (BCGs;
Maurogordato et al.  2008).  The cluster has been observed in the SZE at 30
GHz with OVRO/BIMA (Reese et al.  2002), and at 142, 217, and 268 GHz using
SuZIE (Holzapfel et al.  1997a,b). {Using relativistic corrections to
  the SZE, Hansen et al. (2002) used these data to directly constrain the mean
  ICM gas temperature in Abell 2163, albeit with very large uncertainties.}

During the commissioning observation of
the APEX-SZ bolometer camera (Schwan et al. 2003, Dobbs et al.  2006) in 2007,
we detected Abell 2163 with high significance ($12\sigma$) at 150 GHz with
arcminute resolution.  We observed the cluster in the SZE increment at 345 GHz
using the LABOCA bolometer camera (Kreysa et al. 2003, Siringo et al. 2009) on
APEX, providing images with an angular resolution of 19.5$^{\prime\prime}$.

With a map larger than $12^{\prime}$ across, the present LABOCA data on Abell
2163 provide the first large-area imaging of a galaxy cluster at sub-mm
wavelengths.  High-resolution SZE increment imaging is important in view of
planned future observations of clusters at sub-mm wavelengths, and
for controlling potential foregrounds. However, due to limitations in
constraining the SZ emission at large radii, the use of the sub-mm data in
this paper is limited to the application of the multi-frequency SZE
measurement to model the SZE spectrum and derive constraints on the thermal
and kinematic SZE components.

In sections \ref{sec:obs} and \ref{sec:red} we describe the observations and
data reduction algorithms used to obtain maps at 150 and 345 GHz. We also
describe the analysis of complementary XMM-Newton data. In section
\ref{sec:iso}, a simple isothermal model of the cluster gas density
distribution is presented. A joint SZE/X-ray de-projection of the cluster
temperature and density structure is performed in section \ref{sec:nonisot}.
Finally, we incorporate all available SZE observations of Abell 2163 in
section~\ref{sec:spectrum} to derive the best-fit line-of-sight peculiar
velocity and central Comptonization from the SZE spectrum.  We list our
conclusions in section~\ref{sec:concl}.


\section{Observations}
\label{sec:obs}

\subsection{APEX-SZ observations at 150 GHz}

Observations of Abell 2163 at 150 GHz were carried out in April 2007 with the
APEX-SZ bolometer camera, with a measured beam full-width half-maximum (FWHM)
of 58$^{\prime\prime}$ (determined from observations of Mars) and an effective
bandwidth of 23 GHz (Halverson et al.  2008, hereafter H09). At the time of
observation, 280 of the 330 bolometers were read out, and of these about 160
were used for the map making after applying a noise cutoff (section
\ref{sec:red}).

To obtain a uniform coverage over the extended emission of the cluster,
horizontal raster scans were made with a $15' \times 15'$ scan pattern,
tracking the central cluster position. The median scan speed was
230$^{\prime\prime}\, $s$^{-1}$, and the total integration time was 7.6 hours.
The observing conditions were good to moderate for the site, with 150 GHz
zenith opacities in the range $0.04$-$0.09$.

To test a different scan strategy, 0.8 hours of additional observations were
carried out in August 2007, scanning in circles of radius 6 arcminutes. At the
beginning of each 2-minute subscan, the circle center was placed on the
cluster center, and during the subscan the source was allowed to drift through
the scan pattern.

\subsection{LABOCA observations at 345 GHz}

After producing a high signal-to-noise detection with APEX-SZ, five hours of
follow-up observations at 345 GHz with LABOCA were completed in September
2007. Seven further hours of observations at this frequency were carried out
in May 2008.

At the time of observation, LABOCA had 268 optically active bolometer
channels, of which around 240 were used for mapping of Abell 2163.  From
observations of Neptune the co-added beam size was determined to be
19.5$^{\prime\prime}$ (FWHM). The observing conditions were stable in both
observing periods, with a typical atmospheric zenith opacity of $0.2$-$0.3$ at
the observing frequency.  Abell 2163 was observed in an elevation range of
$40$-$70$ degrees.

For the LABOCA observations, a spiral pattern was used with inner and outer
radii of 120 and 180 arcseconds, respectively. Each scan was made up of four
such spirals, separated by 140 arcseconds in azimuth and elevation, to make
the coverage close to constant over the central part of the cluster.

\subsection{X-ray observations}

X-ray emission originating from clusters provides additional information on
their ICM, and in the interpretation of our results we make use of archival
XMM observations of Abell 2163. The cluster was observed in August 2000 as
part of the guaranteed time from the first year of operation.  Due to the
large size of the target, as compared to the XMM field of view, a mosaic of
five pointings was required to fully cover the cluster: one `on-source'
pointing\footnote{Observation Id: 0112230601} and four offset
pointings\footnote{ObsId: 0112230701, 0112230801, 0112230901, 0112231001} to
probe the outskirts of the cluster and the surrounding cosmic background.  The
nominal exposure time of each observation was about 30ks. Since the full
exposure could not be achieved for the `on-source' pointing, the observation
was completed with a second pointing\footnote{ObsId: 0112231501} in September
2001, which is included in our analysis.


\section{Data reduction, calibration and mapping}
\label{sec:red}

\subsection{General sky-noise considerations}

The SZE signal from Abell 2163 is extended over more than 20 arcminutes, which
is comparable to the field-of-view of APEX-SZ ($23^{\prime}$ across) as well
as that of LABOCA ($11^{\prime}$ across). This makes the subtraction of
atmospheric signals very difficult; subtracting common-mode signals across the
array and applying polynomial baselines (corresponding to large-scale Fourier
modes) in this case has the side effect of also removing astronomical signals
on scales comparable to the field-of-view of the array.

The properties of the two instruments used in this analysis differ
substantially, calling for highly specialized reduction schemes in each case.
In general, the APEX-SZ observations were carried out in poorer observing
conditions than the LABOCA observations, with relatively high levels of
precipitable water vapor. For this reason, the APEX-SZ data suffers from
excess low-frequency noise correlated on scales smaller than the array,
requiring high-pass filtering of individual bolometer time streams to be
applied after removing the correlated atmospheric signal.  While this step
enhances the signal-to-noise ratio of the detection, it also removes
additional astrophysical signal. To account for this, we make use of a point
source transfer function as in H09.  It is used as follows: (i) for the
isothermal analysis, we model the cluster by convolving the parameterized
cluster model with the transfer function and compare the result with our
reduced map; and (ii) for the non-parametric non-isothermal analysis, we
deconvolve the reduced map with the transfer function.

In spite of the higher frequency band, there is no excessive low-frequency
atmospheric noise component in the time streams of the LABOCA data.  However,
because of the lower sensitivity of this instrument with respect to the SZE
signal, a smaller scan pattern was used, compared to the APEX-SZ observations,
to maximize the signal-to-noise ratio in the limited time available. The small
scan pattern fundamentally limits the scale on which the SZE signal can be
recovered. In addition, the details of the LABOCA data reduction limits the
use of a transfer function since the latter cannot be constructed to be
linear; in other words, the reduction pipeline acts differently on a point
source and an extended source.  To recover as much of the SZE increment signal
as possible, we use a slightly modified form of the iterative map-making
algorithm outlined by Enoch et al.  (2006).

For the complete reduction, the bolometer array data analysis software 
BoA\footnote{http://www.apex-telescope.org/bolometer/laboca/boa/}
has been used.

\subsection{Calibration of the millimeter data}

{Primary flux calibration for APEX-SZ and LABOCA relies on observations
  of planets. The APEX-SZ data is calibrated with daily raster scans of Mars
  on all the 280 elements of the array (cf. H09). From these observations the
  gain, beam shape and position on the sky are determined for each bolometer.
  After adjusting the bolometer relative calibrations and angular offsets
  derived from the Mars observations, a co-added rms weighted map is made
  using all optically active bolometers. This map is used to determine the
  overall calibration factor, taking into account the significant side lobes
  of the beam.}

{The calibration procedure is similar for LABOCA, but here the primary
  calibrator is Neptune, which remains unresolved in the 19.5$^{\prime
    \prime}$ beam.  Because Neptune was not always observable, several
  secondary calibrators were used for LABOCA (Siringo et al. 2009). These
  sources used have been monitored by the LABOCA team since early 2007, and
  their 345 GHz flux densities, relative to Neptune, are known to within about
  6\%.}

{The absolute flux of Mars at the APEX-SZ frequency at each observing
  period is determined using a modification of the Rudy model (Rudy et al.
  1987; Muhleman \& Berge 1991), maintained by Bryan
  Butler\footnote{http://www.aoc.nrao.edu/~bbutler/work/mars/model} and
  corrected to be in agreement with recent WMAP results as described in detail
  in H09. The absolute flux of Neptune at 345 GHz is derived from
  cross-calibrations of Neptune against Mars (Griffin \& Orton 1993).}

To monitor the stability of opacity corrections at different elevations,
secondary-calibrator measurements are used. The computation of opacities
relies on skydip measurements combined with tau-meter readings (Weiss et al.
2008, Siringo et al. 2009). 

\subsection{APEX-SZ data reduction}
\label{sec:red:apexsz}

Time stream data from APEX-SZ is processed through a data reduction pipeline
and binned to form maps. This process is described here.  The pipeline is
slightly different for raster scans and circular drift scans, but identical
for each scan within these two subsets.

After eliminating detectors with low optical response, correlated (atmospheric
and electronic) noise is removed by subtracting the median signal from across
the good channels of the array at each time sample.  Individual time streams
are then despiked by flagging and removing data that deviates from the
baseline by more than 5$\sigma$, which typically corresponds to less than $0.1
\%$ of the data. Electronic glitches are recognizable because they occur with
durations shorter than the detector response time. These features, seen only
in a negligible amount of data, are removed as well.

To baseline the data of a raster scan, the scan is divided into subscans of
constant elevation, resulting in subscans extending over 30$^{\prime}$ in
azimuth.  Turnarounds in the scan pattern, where the scan speed is
significantly lower, are flagged and removed. A fifth order polynomial is
fitted to and subtracted from each constant-elevation subscan, effectively
high-pass filtering the time streams.

Circular drift scans are baselined by defining subscans consisting of 2.5 full
circles and applying a fifth order polynomial, after flagging the first full
circle of the scan (which usually contains intervals of high scan
acceleration).  This polynomial baseline corresponds to a spatial filter
similar to that applied to the raster scans.  Simultaneously, an airmass
correction is applied as described in H09.

For each scan, a map with $10^{\prime\prime} \times 10^{\prime\prime}$ sized
pixels is constructed, weighting the data by the inverse rms of each reduced
time stream. All maps are then co-added. In parallel with the reduction of the
data, a beam-shaped source, without noise and translated into time stream
data, is passed through an identical pipeline to obtain the point source
transfer function, as described by H09. All flags and weights on the data are
carried over to the artificial data.

\begin{figure}[H]
\includegraphics[bb=80 10 400 360,width=9cm]{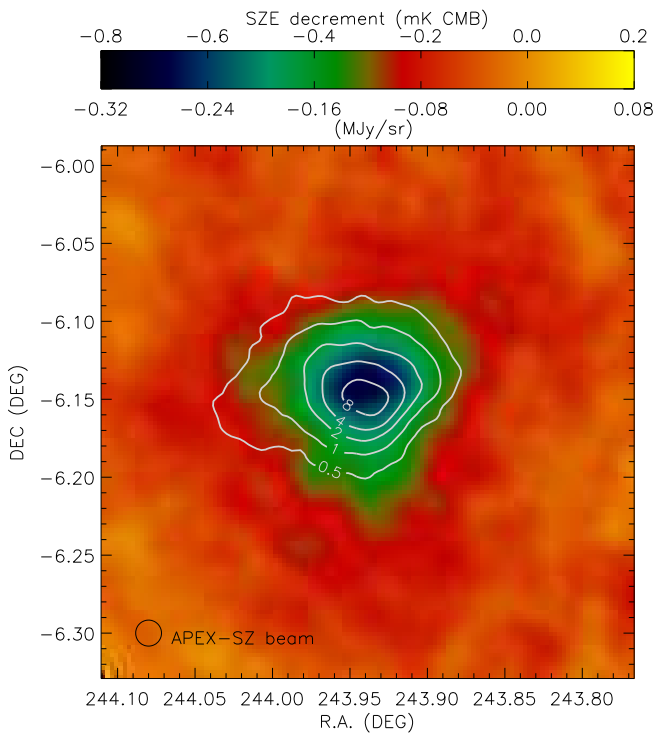}
\caption{{Map of Abell 2163 at 150 GHz}, overlaid with XMM-Newton X-ray
  contours (see Fig. \ref{fig:xmm.map}) in units of $10^{-13}
  \mathrm{erg\,s^{-1}cm^{-2}arcmin^{-2}}$.  Because the correlated-noise
  removal has attenuated the source signal, the map has been de-convolved
  using the point source transfer function (see text). }
\label{fig:aszca.r.m}
\end{figure}

The resulting transfer function is used to deconvolve the high significance
co-added map of Abell 2163 to the intrinsic resolution of the instrument. The
deconvolution is done iteratively, and is similar to the CLEAN algorithm for
interferometry data (H\"{o}gbom 1974, Schwarz 1978) in that the source is
modeled as a sum of many point sources; here, however, the process is carried
out in map space rather than Fourier space. The raw (pipeline filtered) map,
$M_0$, is first \textit{convolved} with the beam to reduce the sample variance
on small spatial scales. Each pixel in the map is then divided by the local
rms to create a signal-to-noise map, $N_0$, from which the brightest pixel is
selected. The corresponding point source flux corresponding to this peak pixel
value is computed, taking into account both the beam smoothing of $M_0$ and
the complete description of the inherent resolution and the effects of the
reduction described by the point source transfer function. The resulting point
source is re-convolved with the transfer function to represent how the point
source is reflected in the raw map. This component is subtracted from $M_0$ to
yield $M_1$, from which a new smoothed signal-to-noise map $N_1$ is
constructed. The point source is also convolved with the ``clean'' beam and
added to a map $C$ of ``clean component''. Next, the brightest pixel in $N_1$
is located, and the process is repeated until $M_i$ is consistent with noise.
The criterion for the latter is that as many negative as positive pixels are
selected in any consecutive 20 iterations.  At this point, $M_i$ is added to
the cumulative map $C$ to account for emission which has not been successfully
removed as point sources. Details on the limitations of this deconvolution
technique will be given in a future publication (Nord et al., in preparation).
The de-convolved APEX-SZ map of Abell 2163 is shown in Figure
\ref{fig:aszca.r.m}.

To compute the noise on the beam scale, we smooth the final map with the
APEX-SZ beam (FWHM 58$^{\prime\prime}$ and computing the rms in square regions
with side five times the beam. In this way, we find a noise rms value of 32
$\mu$K$_{\text{CMB}}$ on the beam scale, corresponding to 0.013 MJy sr$^{-1}$,
in the central region of the de-convolved map.

\subsection{LABOCA data reduction}
\label{sec:red:laboca}

Because the LABOCA time streams are more stable at low frequencies than those
from APEX-SZ, the data do not have to be further filtered after standard
correlated-noise removal. Aside from this difference, the time stream
reduction sequence used in each iteration is quite similar to that used for
the APEX-SZ data, and basically follows the method outlined by Weiss et al.
(2008) and detailed by Siringo et al. (2009). Maps are constructed with
$6^{\prime\prime} \times 6^{\prime\prime}$ pixels, and co-added in the same
way as for the APEX-SZ data.

To account for sky signal attenuation in the time stream reduction, we follow
the iterative approach of Enoch et al. (2006), with one important
modification.  Using a suitably chosen time stream reduction algorithm, a map
is produced.  Each pixel with a significance of less than 3 sigma is set to
zero, and the rest of the map is chosen as the template to be subtracted
directly from the time streams before running the same time stream reduction
yet again, and so on.  While Enoch et al. take a conservative approach and
derive the next template from the residual signal, we add the template back
into the data prior to mapping and derive this template from the total map.
This ensures that any significant feature in the \textit{current} best guess
of the source flux distribution is carried over to the next iteration step.

The iterative mapping algorithm requires 8 iterations before convergence is
reached. In each iteration, an identical time series reduction is performed.

Because the iterative mapping technique is non-linear, it is not
straightforward to characterize it by a filter function. In place of a
transfer function, the instrument beam is used when fitting a model to the 345
GHz data. Significant loss of signal is expected using the iterative approach.
To quantify the level of bias, we use the best-fit $\beta$ model from the
APEX-SZ measurement (section \ref{sec:iso}) and pass is through the complete
LABOCA data reduction pipeline. Although more than $30\%$ of the signal is
lost beyond $r_{500}$ (the radius at which the enclosed average matter
overdensity is 500 times the mean cosmic density), we find that
the signal loss within $r_{2500}$ (defined analogously, approximately $3^{\prime}$) is negligible compared
to statistical errors. The central rms in the LABOCA map, computed on a scale
of five times the LABOCA FWHM (19.5$^{\prime\prime}$), is 1.9 mJy/beam,
corresponding to 0.24 MJy sr$^{-1}$.

\begin{figure}[H]
\centering
\includegraphics[width=9cm]{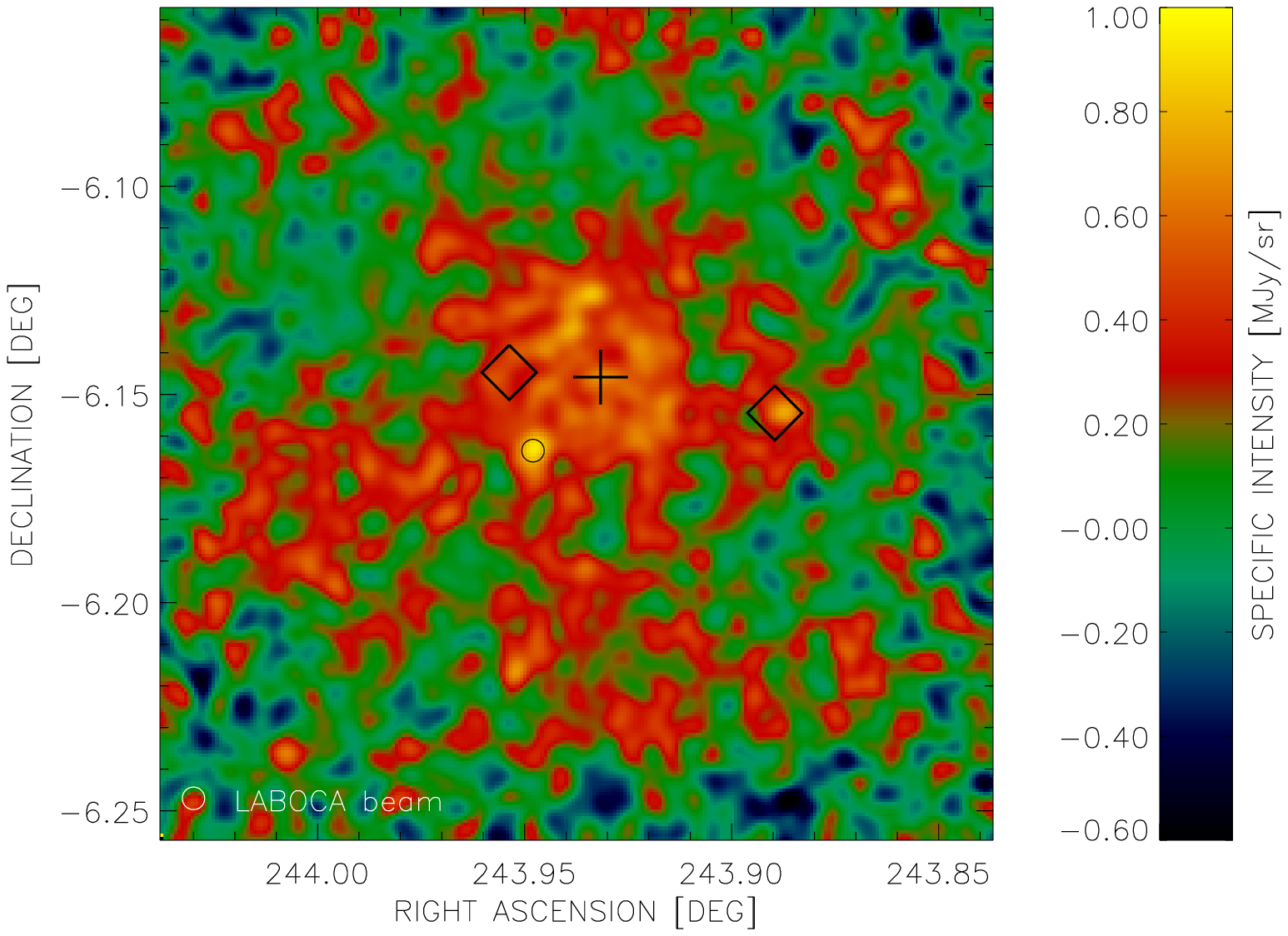}
\includegraphics[width=9cm]{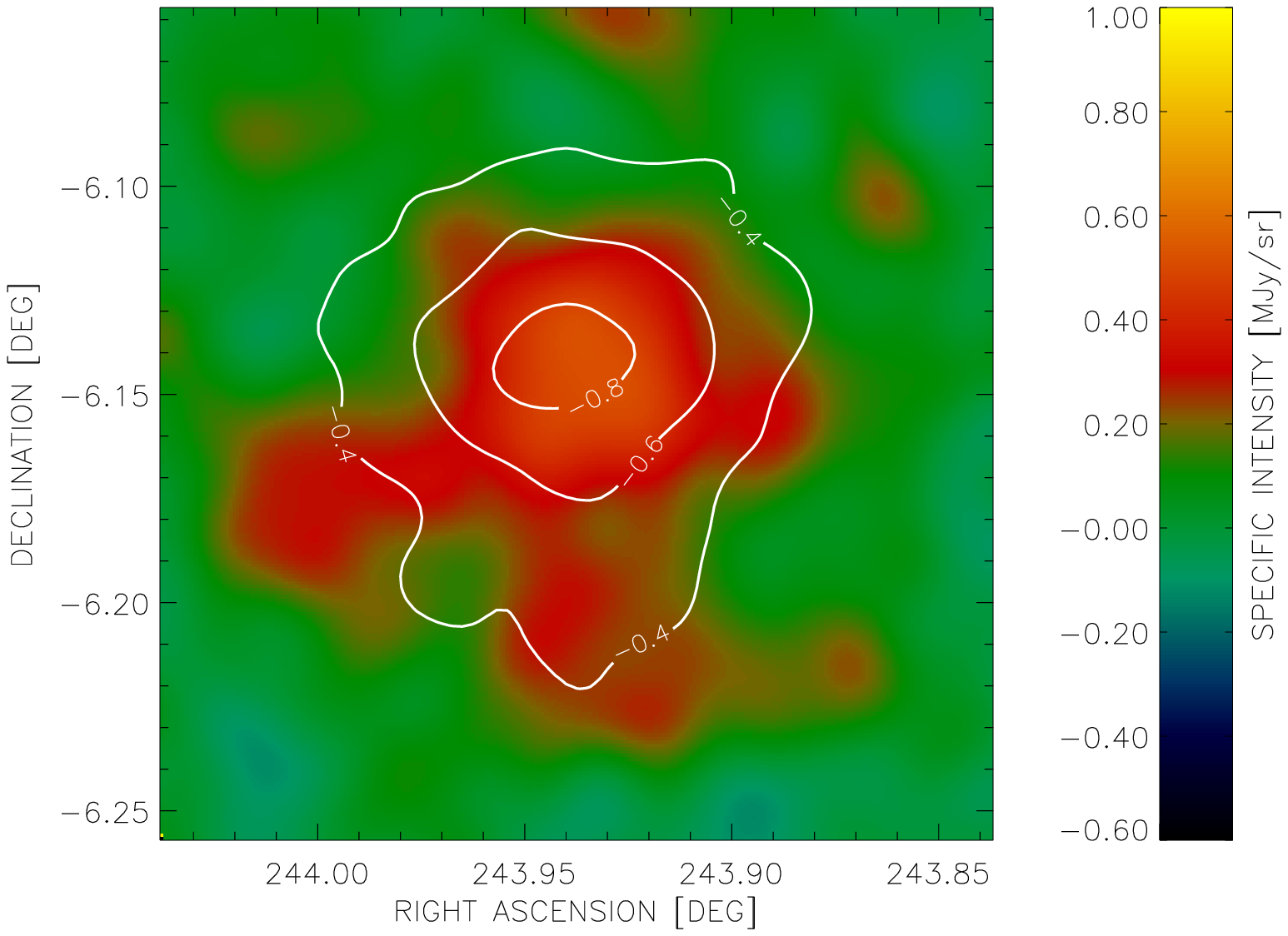}
\caption{\textit{Top:} Final 345 GHz LABOCA map of Abell 2163, smoothed with
  the 19.5$^{\prime\prime}$ beam.  The cross marks the position of a bright
  flat-spectrum radio source (Cooray et al. 1998). The diamonds mark the
  positions of two BCGs (Maurogordato et al. 2008). The circle marks the
  position of the bright point source found in the present data.
  \textit{Bottom:} LABOCA map smoothed to the APEX-SZ resolution of 1
  arcminute. The bright point source described in the text has been removed.
  The APEX-SZ 150 GHz map is shown as contours.}
\label{fig:labocamaps}
\end{figure}

We find one significant point source in the final map (Figure
\ref{fig:labocamaps}), likely thermal emission from a cluster galaxy or a
background source lensed by the cluster. After highpass-filtering the map to
remove structures larger than two beams (such as the cluster emission), we
find a peak flux density of $11.8 \pm 1.9$ mJy/beam for this source.

\subsection{XMM-Newton data reduction}
\label{sec:xreduc}

Here we describe the processing steps applied to the X-ray data.

We generate calibrated event-lists from the raw data of the 6 archival 
pointings using the standard procedures of the XMM Science Analysis 
System\footnote{http://xmm.esac.esa.int/sas/, all the data processing for this paper relies on v7.1.2} 
(SAS) and screen these lists for periods of high radiation due to soft solar proton flares.  
To this end, we apply the method of Pratt \& Arnaud (2002), which consists of fitting 
histograms of the high energy\footnote{more precisely [10-12]~keV for the MOS detectors and 
[12-14]~keV for the PN} light curves with a Poisson law and rejecting time intervals
exceeding the mean radiation level by more than 3$\sigma$. On average, this
shortens the available exposure time by 15\%.  Although the high energy signal
is very sensitive to particle flares, one needs to adopt a coarse time
sampling (104s here) to ensure that the average number of counts per bin is
significant. This implies that the wings of light curve jumps in increased
radiation periods cannot easily be detected.  To improve on our first
filtering, we thus repeat the same analysis on a broad low energy band, namely
[0.3-10]~keV with narrow time bins of 13s, resulting in the exclusion of an
extra 3\% of the data. {Out-of-time event lists are also generated using the 
simulation mode of the SAS tasks emchain/epchain and filtered using the same  
time intervals.}

From the filtered event lists, images and exposure maps are generated in the
[0.5-2]~keV band and gathered into two large mosaics. {The signal to
  noise of the cluster in this band is about 80\% of the best achievable value
  using the XMM bands. It is however free from the strongly varying
  instrumental lines around 6-7 keV and calibration uncertainties at lower
  energies. Combined with the weak dependence of the cooling function on
  temperature at low energies (soft X-rays), this makes it a very suitable
  band for our purpose. Since the XMM response is almost flat in this narrow 
  band, the exposure maps are evaluated at a single average energy of 1.25~keV.}

{The main issue in the X-ray analysis of extended source is an accurate 
background modeling. To this end, we make use of the Filter Wheel Closed 
(FWC) dataset provided by the XMM background working 
group\footnote{http://xmm2.esac.esa.int/external/xmm\_sw\_cal/background/}. 
Our model consists of a sum of a re-scaled FWC image (which represents the 
instrumental noise), an homogeneous cosmic background of constant surface 
brightness (since the area of the mosaic remains small) and a properly scaled 
image of the out-of-field events. 
The scaling of the instrumental background is allowed to vary with each 
pointing and instrument while the cosmic background level is just a function 
of the detector (to account for the different responses). Using the Cash 
statistics (Cash 1979), the model parameters are fitted simultaneously to the 
6 observations including the data from the out of field corners and excluding 
the positions closer than 17$^{\prime}$ from the cluster center.}

{In order to detect and mask the surrounding AGNs, we adaptively smooth the 
background subtracted raw mosaic, using the criterion of having more than 
10 counts per cell. The resulting map is then exposure corrected and smoothed
again by a $\sigma=2^{\prime\prime}$ Gaussian\footnote{This hybrid smoothing 
allows to wash out the Poisson fluctuations in regions devoid of source signal 
without degrading the resolution over moderately bright sources (which are 
both the advantage and drawback of adaptive filtering alone)}.
A source catalog is extracted from this image using SExtractor (Bertin \& Arnouts 1996)
and the associated source-segmentation map is combined with an exposure time 
threshold in order to derive a global mask.} 

{The tasks of background estimation and source detection are somewhat
  intricate, especially in the case of mosaics where steep instrumental and
  particle background variations can occur in between pointings. To tackle
  this issue, we iteratively re-perform the background modeling and AGN
  extraction using the updated AGN mask for the background estimate. Both
  products stabilize after three iterations.}

\begin{figure}
\includegraphics[width=9cm]{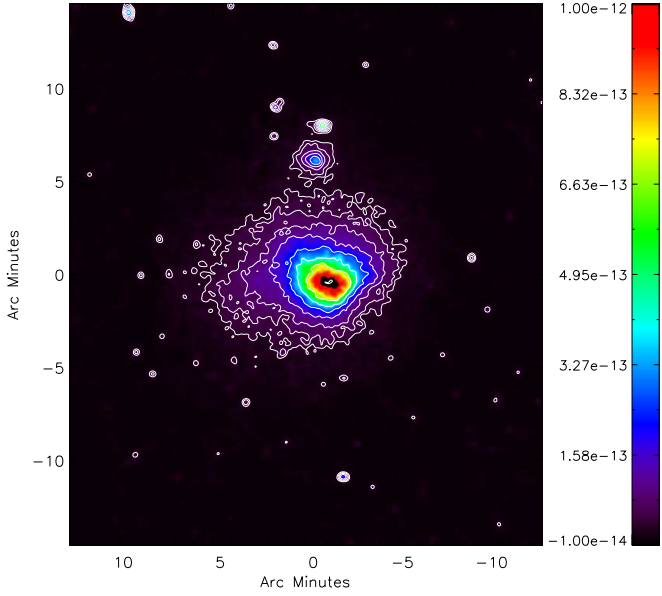}
\caption{Smoothed, background subtracted, X-ray map of Abell 2163 in the
  [0.5-2]~keV band (see text for details). Logarithmically spaced contours
  highlight the broad dynamical range of the cluster emission. The unit of the
  color scale is $\mathrm{erg\,s^{-1}cm^{-2}arcmin^{-2}}$. }
\label{fig:xmm.map}
\end{figure}

The final background subtracted, adaptively smoothed image of the cluster is 
shown in Fig.~\ref{fig:xmm.map}.

\section{Isothermal modeling of the intra-cluster gas}
\label{sec:iso}

In order to assess our SZE measurements quantitatively and compare them with
the X-ray data, we fit the SZE measurements using the well-known isothermal
$\beta$ model (Cavaliere \& Fusco-Femiano 1978). This enables us to compare
the X-ray surface brightness profiles with the SZE temperature
decrement/increment maps and check the agreement of our SZE measurements with
the prediction from the X-ray mean values.

\subsection{Isothermal gas in hydrostatic equilibrium}

The isothermal $\beta$ model is widely used to model the gas profile in
clusters {under the assumption of hydrostatic equilibrium}.  The radial
distribution of gas density is given by
\begin{equation}
n_e(r) = n_e(0) \left(1+ \frac{r^2}{r_c^2}\right)^{-\frac{3}{2}\beta},
\label{eq:beta}
\end{equation} 
where $n_e$ is the density of electrons, $r$ is the radius from the center,
$r_c$ is the core radius of the gas, and $\beta$ is the power law index.

With this model, the temperature decrement (or increment) due to the SZE takes the
form
\begin{equation}
 \Delta T(\theta) = f(x,T_e) \ T_{\mathrm{CMB}} \ y 
  = \Delta T(\theta=0)
  \left(1+\dfrac{\theta^2}{\theta_c^2}\right)^{(1-3\beta)/2},
\label{eq:szprof}
\end{equation} 
where $\theta$ is the angular radius on the sky, $\Delta T$ is the
\textit{thermodynamic} SZE temperature decrement (or increment)
and $y$ is the Comptonization parameter, given by the line-of-sight
integral
\begin{equation}
\label{eq:compton}
y = \int \left(\frac{k_{\mathrm{B}}T_e}{m_ec^2}\right) \ n_e \sigma_{\mathrm{T}} \ \mathrm{d}l,
\end{equation}
where $\sigma_{\mathrm{T}}$ is the cross section for Thomson
scattering. The frequency dependence $f(x,T_e)$ of the SZE signal is given by
\begin{equation*}
 f(x,T_e)=\left[x \frac{e^x+1}{e^x-1} -4\right](1+\delta_{x,T_e}),
\end{equation*}
with relativistic
corrections contained in $\delta_{x,T_e}$ as described by Sazonov \& Sunyaev
(1998).
$x$ is the dimensionless frequency $x = \frac{h \nu}{k_B T_0}$, 

The analogous expression for the X-ray surface brightness reads
\begin{equation}
  \Delta S_X(\theta) = S_0 \left(1+\dfrac{\theta^2}{\theta_c^2}\right)^{1/2-3\beta},
\label{eq:xprof}
\end{equation}
where $S_0$ is the central surface brightness and the different exponent
comes from the $n_e^2$ dependence of X-ray emissivity.

\subsection{Elliptical $\beta$ model fit to the SZE data}
\label{sec:beta:sz}

Because of the significant ellipticity in the X-ray surface brightness
profile, we generalize equation \ref{eq:szprof} to an elliptical form given,
e.g., in H09 as
\begin{equation}
\label{EQN:ellipticalbeta}
\Delta T_{\rm{SZE}} = \Delta T_{0}\left(1 + A + B \right)^{(1-3\beta)/2},
\end{equation}
where
\begin{equation*} 
A = \frac{(\cos(\Phi)(X - X_{\rm{0}}) + \sin(\Phi)(Y -
  Y_{\rm{0}}))^2}{\theta_{\rm{c}}^2},
\end{equation*}
\begin{equation*} 
B = \frac{(-\sin(\Phi)(X - X_{\rm{0}}) + \cos(\Phi)(Y - 
  Y_{\rm{0}}))^2}{(\eta \theta_{\rm{c}})^2}.  
\end{equation*}
Here $(X - X_{\rm{0}})$ and $(Y - Y_{\rm{0}})$ are angular offsets on the sky
in the right ascension (RA) and declination (DEC) directions, respectively,
with respect to centroid position ($X_{\rm{0}}$, $Y_{\rm{0}}$). The axial
ratio, $\eta$, is the ratio between the minor and major axis core radii and
$\Phi$ is the angle between the major axis and the RA ($X$) direction.

An elliptical $\beta$ profile is fitted to the 150 GHz data by convolving each
model function with the point source transfer function and minimizing a
$\chi^2$ statistic weighted by the inverse square of the local map rms. Note
that the raw, not the de-convolved, map is used for this fit in conjunction
with the transfer function. All map pixels within a 10$^{\prime}$ radius of
the X-ray center (section \ref{sec:beta:x}) are considered for the fit. The
results of the fit are given in Table \ref{tab:betafits}.

For comparison, we also fit a spherical $\beta$ profile and find roughly
consistent results. At 345 GHz, the detection of Abell 2163 is less
significant, we thus fix the centroid and the value of $\beta$ to those fitted
from the 150 GHz measurement and fit for the remaining parameters.

Errors in the fitted profiles are estimated using a Monte Carlo approach. The
map data is jack-knifed by inverting the sign of half the individual scan maps
(randomly chosen), and the maps are co-added to generate a pure noise map.
After convolving the best-fit model with the transfer function and adding the
result to the noise map, the $\chi^2$ statistic is again minimized to find the
best set of parameters for the new map. The entire procedure is repeated 1000
times, upon which the errors in the respective parameters are taken as the
scatter in the distributions of their fitted values. We verify that the mean
of the distributions of fitted parameters are consistent with the best-fit
model. 

\subsection{Elliptical $\beta$-model fit to the X-ray data}
\label{sec:beta:x}

To allow for a basic comparison of the SZE and X-ray data in the context of the
isothermal $\beta$ model, we perform an isothermal fit to the X-ray data,
similar to that used for the SZE data.  As for the SZE analysis, we generalize
the spherical $\beta$ model of equation~\ref{eq:xprof} to the elliptical case, as:
\begin{equation}
\label{EQN:Xellipticalbeta}
\Delta S_{\rm{X}} = \Delta S_{0}\left(1 + A + B \right)^{1/2-3\beta},
\end{equation}

This model is fitted to the raw X-ray images using the Cash statistic (Cash
1979).The Poisson nature of the data does not allow here for a direct
background subtraction; instead we add the simple two-component background
described in section \ref{sec:xreduc} to the fitted components. The results of
the fit are presented in Table \ref{tab:betafits}. 

\subsection{Systematic uncertainties}
\label{sec:iso:sys}

\subsubsection{Galactic dust}

The effect of galactic dust emission on the SZE measurement of Abell 2163 has
been estimated by Lamarre et al. (1998) for the PRONAOS balloon experiment and
by LaRoque et al. (2002) for the SuZIE experiment (Holzapfel et al. (1997a,b)).
However, because these two estimates apply to scan strategies and spatial
filtering functions very different from ours, it is not appropriate to
extrapolate from either of them.

To estimate the dust correction at 345 and 150 GHz, we use the IRAS
far-infrared $60\,\mu$m and $100\,\mu$m dust maps (Schlegel et al. 1998) on the
Abell 2163 field . Smoothing the $60\,\mu$m map to the $4.3^{\prime}$
resolution of the $100\,\mu$m map and re-sampling the maps so that each pixel
has a side of one arcminute, we extrapolate to the frequencies of the SZE
measurements using a grey-body spectrum of dust emissivity
\begin{equation}
F_d \propto \frac{\nu^{3+\alpha}}{\exp(h\nu/k_{\mathrm{B}}T_d) - 1},
\label{eq:galdust}
\end{equation}
where $\alpha$ is the dust spectral index. Using $\alpha=2.0$ (Finkbeiner et
al. 1999) we fit to $T_d$ for every $1^{\prime}\times1^{\prime}$ pixel pair in
the two dust maps. After smoothing the resulting temperature map with a
$4.3^{\prime}$ Gaussian kernel to remove noise artifacts, we use equation
(\ref{eq:galdust}) to extrapolate the intensity at 345 and 150 GHz,
normalizing at $100\,\mu$m.

The simplest way of removing the dust component from the data is to subtract
it before any filtering is applied by the reduction. We thus create time
streams corresponding to the extrapolated dust maps on the coordinates covered
by APEX-SZ and LABOCA and subtract these from the raw time streams before
proceeding with reductions identical to those described in section
\ref{sec:red}. In this way, we find corrections of -0.6$\%$ to the SZE
decrement at 150 GHz and +1.8$\%$ to the SZE increment at 345 GHz. 
Effects on the other fitted parameters are negligible. We assume that
contamination on angular scales smaller than four arcminutes (i.e. not
resolved by IRAS) are also negligible.

\subsubsection{Point source contamination}

After subtracting a preliminary (possibly point source contaminated) $\beta$
model from each map, we smooth the maps to the respective beam scales and look
for point source detections above $3 \sigma$. One source (cf. Figure
\ref{fig:labocamaps}) is found in the LABOCA map at 345 GHz. No sources are
found in the APEX-SZ 150 GHz map. Point sources which are detected with
sufficient S/N in the maps could be removed by subtracting fitted Gaussians.
Here we take a somewhat more conservative approach and simply disregard a
region around the 345 GHz point source in a subsequent fit with a new $\beta$
model.

Extragalactic infrared point sources are not expected to correlate spatially
with clusters of galaxies, and a distribution of sources can be well
approximated by excess noise in any differential measurement (White \&
Majumdar 2004), thus having little systematic effect on bolometric SZE
measurements.  

Although lensing by a massive cluster such as Abell 2163 can significantly
increase the number of detected sources in the cluster central region, this
effect conserves the total intensity of the background and only becomes
important when faint sources are raised above the detection limit and removed,
and the unresolved background intensity is thereby systematically lowered
(e.g. Loeb \& Refregier 1997).  It is not clear whether the bright point
source found in the 345 GHz map is a lensed background source; however, we
have verified that the effect in fitted parameters of excluding the region
around this source for the fit is merely 4\%, i.e. smaller than both the
derived statistical errors and the absolute calibration uncertainty.

Radio sources are expected to correlate strongly with clusters of galaxies
(e.g. Reddy \& Yun 2004). We have searched the NVSS catalog (Condon et al.
1998) for sources brighter than 5 mJy at 1.4 GHz.  The flux of each source is
extrapolated to 150 GHz using the spectral index maps of Feretti et al.
(2004). With this simple approach, even the brightest source will have a peak
flux considerably lower than the rms in the APEX-SZ map. Cooray et al.  (1998)
report an inverted-spectrum non-thermal source of 3 mJy at 30 GHz, indicated
in Figure \ref{fig:labocamaps}. No counterpart can be seen in either SZE map.
High-pass filtering the APEX-SZ map, after subtracting the best-fit model of
the cluster, to remove all structure larger than $3^{\prime}$ results in an
effective rms of 30 $\mu \text{K}_{\text{CMB}}$ at the position of this source. The
fact that the source is not seen in this filtered map allows us to put a lower
limit on the spectral index as $\alpha_{30}^{150} \gtrsim 0.5$ ($S_{\nu} \propto
\nu^{-\alpha}$). Subtracting the corresponding signal from the APEX-SZ map
before the $\beta$ model fit results in a systematic shift in the central SZE
decrement of less than 1\%. Based on these considerations, we consider
systematic effects from radio sources negligible in comparison to primary CMB
anisotropies and absolute calibration.

\subsubsection{Primary CMB contamination}

For a very extended cluster like Abell 2163, with an estimated $r_{500}$
greater than $7^{\prime}$ (Radovich et al. 2008), the temperature anisotropies
in the CMB are a potential source of confusion in the SZE measurements. This
uncertainty particularly affects the kinematic SZE signal because the latter
has a frequency dependence identical to that of the CMB thermal spectrum. Note
that this is not an issue for CMB power on scales $\gtrsim 15^{\prime}$ due to
the indirect high-pass filtering of our maps (through polynomial baselines in
the bolometer time series).

To quantify the level of contamination on scales comparable to the SZE
emission, the HEALpix software (Gorski et al.  2005) is used to generate 100
realizations of the CMB sky in a one square degree field with a resolution of
1.7$^{\prime}$. The input power spectrum for this generation is computed using
the CMBfast code (Zaldarriaga \& Seljak 2000) with the WMAP 5-year cosmology
(Komatsu et al. 2009). Relying on the results of LaRoque et al. (2002),
diffuse secondary CMB anisotropies other than the SZE are considered
negligible.

Each primary-CMB map is re-sampled as time streams corresponding to a typical
scan of the SZE observations at the two frequencies and run through the
reduction pipelines to account for filtering. The $\beta$ model fit is repeated
after subtracting each of the thus filtered primary-CMB realizations.  The scatter
in the results of each parameter is taken as the 68\% systematic marginal
error from primary CMB contamination. We find a fractional error of $3.4\%$ at
150 GHz and $2.9\%$ at 345 GHz.

The primary-CMB component has been simulated on the relevant scales compared
to the cluster emission, with scales smaller than 1.7$^{\prime}$ not taken
into account.  This is justified because the power of the primary anisotropies
drops rapidly on scales smaller than a few arcminutes.

\subsubsection{Pipeline filter function}
\label{sec:iso:sys:pipe}

To test how well the point source transfer function represents the filtering
of the SZE signal in map space at 150 GHz, we carry out a series of reductions
of known signals. To this end, we remove astrophysical signals from the data
by dividing the scans into two minute segments and randomly inverting half of
these.  Before passing the noise data through the pipeline, a beam-convolved
$\beta$ model is added to the time streams. The point source transfer function
is used to reconstruct the $\beta$ model after reduction, and the procedure is
repeated for a large number of $\beta$ models in the range $50^{\prime\prime}
\leq \theta_c \leq 150^{\prime\prime}$, $0.5 \leq \beta \leq 1.4$. We find no
evidence that the level of noise affects the result other than to increase the
scatter in fitted parameters; we thus use bright model sources ($\sim10 \,
\text{mK}_{\text{CMB}}$ at 150 GHz) to estimate the effect of the pipeline.

Although the fitted central amplitude can be systematically reduced by as much
as $12 \%$ for combinations of small $\beta$ and large $\theta_c$
(characterizing very broad profiles, where some modes in the map are
irrecoverable by the transfer function), we find the systematic shift to be no
more than $2\%$ for either of the models fitted to the actual 150 GHz data.
Due to the degeneracy between $\beta$ and $\theta_c$, there is considerable
scatter in these parameters when fitted to the simulated data. Keeping one
parameter fixed and fitting the other, we find that the systematic errors in
both parameters due to inaccuracies in modeling the pipeline filter function
are negligible.

\subsection{Results of the isothermal modeling}

\begin{figure}
\centering
\includegraphics[width=9.1cm]{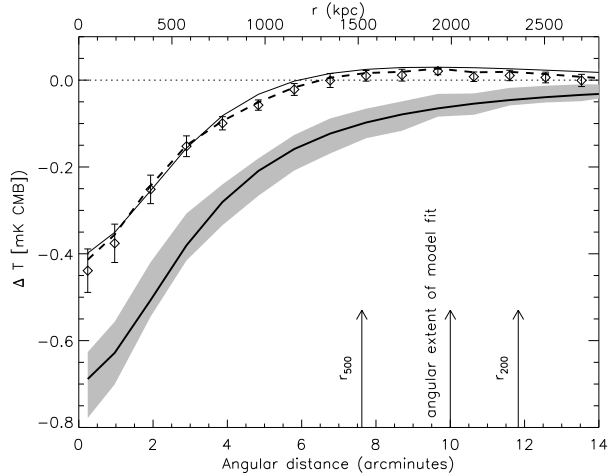}
\caption{{Radial SZE profile of Abell 2163 at 150 GHz. Error bars}
  indicate the profile computed from the reduction-attenuated raw map (before
  deconvolution; this is the map used for the parametric fit) while
  {the shaded region} represents the profile from the map de-convolved
  to the beam resolution (note that this map {was \textit{not} used to
    derive the best fit $\beta$ model.} 
  {The profile of the best-fit isothermal $\beta$
    model is indicated by solid lines; the thick line represents that model
    convolved with the \textit{beam}, while the thin line represents the same
    model convolved with the \textit{transfer function}. The dashed line
    indicates the profile of the de-convolved map, directly re-convolved with
    the transfer function.}
The vertical
  arrows indicates the radial cut of 10$^{\prime}$ used for the model fit, as
  well as $r_{200}$ and $r_{500}$.}
\label{fig:aszca:profile}
\end{figure}

Figure \ref{fig:aszca:profile} shows the radial profile of the best-fit
spherical $\beta$ profile to the 150 GHz data and the same profile convolved
with the transfer function, as well as the profiles of the raw and
de-convolved maps. The best-fit $\beta$ model parameters, corrected for
systematic effects as described below, are given in Table \ref{tab:betafits}.
{Although derived from a completely independent method,} the radial
profile of the de-convolved map
is fully consistent with the beam-smoothed best-fit 
model within the 10$^{\prime}$ truncation radius used for the fit.

Comparing the results of the X-ray and SZE fits, a 
non-negligible offset ($21 \pm 8^{\prime\prime}$), possibly caused by
an asymmetric temperature distribution near the center, appears between the
two centroid positions.  The origin of this offset will be discussed 
as part of further analysis in a future publication (Kneissl et al., in
preparation).  The other fitted parameters show an overall consistency in the
global shape of the emission.  We note that in spite of the significant
ellipticity in the X-ray surface brightness, the elliptical model of the SZE
signal is only barely inconsistent with the spherical one (the axial ratio
deviates from 1 by only 1.7$\sigma$). The most likely explanations for this
are the poorer resolution of the SZE data which results in a greater
uncertainty on the ellipticity, and the inherent fact that the SZE signal, due
to the density weighting, is naturally smoother and more diffuse.

\begin{table*}
\begin{center}
\caption{$\beta$ model fit results at 150 and 345 GHz. }
\label{tab:betafits} 
\centering          
\begin{tabular}{l r@{}l r@{}l r@{}l r@{}l}
\hline\hline   
Parameter & 150 GHz \,& (elliptical) & 150 GHz \,& (spherical) & 345 GHz
\,& (spherical) & X-ray \,& (elliptical) \\ 
\hline

$X_0$ (Central R.A. [J2000]) & 
16$^{\text{h}}$15$^{\text{m}}$45.1$^{\text{s}}$&$\pm 9^{\prime\prime}$ & 
16$^{\text{h}}$15$^{\text{m}}$45.6$^{\text{s}}$&$\pm 8^{\prime\prime}$ & 
(16$^{\text{h}}$15$^{\text{m}}$45.6$^{\text{s}}$)$^b$& 
& 16$^{\text{h}}$15$^{\text{m}}$46.6$^{\text{s}}$&$\pm 1.5^{\prime\prime}$ \\

$Y_0$ (Central DEC. [J2000]) & 
$-$06$^{\circ}$08$^{\prime}$31$^{\prime\prime}$&$\pm 8^{\prime\prime}$ & 
$-$06$^{\circ}$08$^{\prime}$28$^{\prime\prime}$&$\pm 8^{\prime\prime}$ & 
($-$06$^{\circ}$08$^{\prime}$28$^{\prime\prime}$)$^b$ & 
& $-$06$^{\circ}$08$^{\prime}$44.1$^{\prime\prime}$&$\pm 1.4^{\prime\prime}$ \\
        
Peak signal$^a$ &
$-${0.301}&$\pm${0.033} & $-${0.302}&$\pm${0.033} & {0.613}&$\pm${0.089} & 
$1.075 \times 10^{-12}$&$\pm 7 \times 10^{-15}$ \\


$\theta_c$ (core radius) & {123} &$\pm${19}$^{\prime\prime}$ & {129}&$\pm${21}$^{\prime\prime}$ &
{89}&$\pm${29}$^{\prime\prime}$ & 
$90.7$&$\pm 2.8^{\prime\prime}$ \\

$\beta$ (power-law index) & {0.85}&$\pm${0.14} & {0.85}&$\pm${0.14} & ({0.85})$^b$& & 
$0.639$&$\pm 0.03$ \\ 

$\Phi$  (inclination angle) & {128}&$\pm${24}$^{\circ}$ & (0.0)$^b$& & (0.0)$^b$& & 
$1.2$&$\pm 6^{\circ}$\\

$\eta$ (axial ratio) & {0.77}&$\pm${0.11} & (1.0)$^b$& & (1.0)$^b$& & 
$0.850$&$\pm 0.05$\\
\hline
\end{tabular}
\end{center}
$^a$ Brightness 
in units of MJy sr$^{-1}$ (SZE), and in units of
$\mathrm{erg\,s^{-1}\,cm^{-2}\,arcmin^{-2}}$ (X-ray) \\
$^b$ Fixed parameters indicated by parentheses
\end{table*}


\section{Non-isothermal modeling of the intra-cluster gas}
\label{sec:nonisot}

Several authors have discussed the joint modeling of SZE and X-ray data for
de-projecting ICM parameters (see, e.g., Yoshikawa \& Suto 1999, Zaroubi et
al. 2001, Puchwein \& Bartelmann 2007, Ameglio et al. 2007).  However, most of
these analyses have been limited to analytic cluster models or numerical
simulations. Our work here represents the first attempt at de-projecting
cluster density and temperature profiles using actual SZE imaging data, in
combination with the X-ray data, {without resorting to any parametric models 
for the gas or dark matter distribution.}

Given the large angular size of Abell 2163, the arcminute resolution of
APEX-SZ is sufficient to carry out a joint radial density and temperature
modeling of the ICM in combination with publicly available X-ray data.
Although the LABOCA map has more than a factor of two better resolution, due
to its high noise level and small field of view we restrict the analysis
in this section to the APEX-SZ data.

\subsection{Method}

\subsubsection{Deprojection method}

For the deprojection analysis we use a direct inversion technique based on the
Abel integral. The technique was first proposed by Silk \& White (1978; see
also Yoshikawa \& Suto 1999). Apart from the assumption of spherical symmetry,
there are no additional theoretical constraints (e.g. hydrostatic equilibrium
or polytropic gas index), making this is a natural method for extracting ICM
parameters, assuming that spherical symmetry indeed provides a good estimate
of the structure. The details of this method and its limitations will be
discussed in a future publication (Basu et al., \textit{in preparation}); here
we present the outline of the method and the results for Abell 2163.  X-ray
spectroscopic measurements have shown that Abell 2163 has a complicated
temperature structure at the center and is most likely a merging system (Elbaz
et al.  1995, Govoni et al. 2004).  Nevertheless, the de-projection analysis
with spherical symmetry implemented here shows how resolved SZE images of
clusters can immediately be used in combination with X-ray data to gain a
better understanding of the gas and mass distribution.

The deprojection analysis is based entirely on the radial profiles of X-ray
surface brightness and SZE temperature decrement (or increment). From
equation (\ref{eq:szprof}) the SZE temperature decrement can be written as the
integral over the line of sight;
\begin{equation}
\Delta T(R) = 2 A_{\mathrm{SZ}} 
\int_R^{\infty} f(x,T_e)\ n_e(r)\ 
T_e(r)\ \frac{r\mathrm{d}r}{\sqrt{r^2 - R^2}}
\label{eq:szrad}
\end{equation} 
where the cluster is assumed to be spherically symmetric, $A_{\mathrm{SZ}} =
\sigma_{\mathrm{T}}\ (kT_{\mathrm{CMB}}/m_ec^2)$, $r$ is the physical radius
from the cluster center, $R=D_A \theta$ where $\theta$ is the projected
angular distance on the sky and $D_A$ is the angular diameter distance (to
calculate $D_A$, we assume the cosmology of Komatsu et al. 2009 for the
remainder of the paper). $T_e(r)$ and $n_e(r)$ are the electron gas
temperature and density radial profiles. We neglect the small $T_e$ dependence
in $f(x,T_e)$ for this analysis 
and incorporate $f(x,T_e$=$10~$keV$)$ into the $A_{\mathrm{SZ}}$ factor.

For the de-projection analysis, the X-ray surface brightness profile can be
written as (e.g. Yoshikawa \& Suto 1999):
\begin{equation}
S_{\mathrm{X}}(R) = \frac{2}{4\pi (1+z)^4}\ 
\int_R^{\infty} n_e^2(r)\ \Lambda_H(T_e(r))\  \frac{r\mathrm{d}r}{\sqrt{r^2 - R^2}}.
\label{eq:xrad}
\end{equation}

We use the APEC code (Smith et al. 2001) to compute the cooling function,
$\Lambda(T_e(r))$ for Abell 2163, assuming an abundance value of $0.4
Z_{\odot}$, where $Z_{\odot}$ is the solar metallicity. {This also includes 
the absorption by the neutral hydrogen of the Galaxy using a column density 
of $1.1\times10^{21}~{\rm cm}^{-2}$, as measured by the L.A.B. survey (Kalberla et al. 2005) at the cluster position.
While the temperature dependence of the cooling function is thus included in
  the analysis, it is interesting to note that the results changes very little
  when assuming $\Lambda$ is independent of $T_e$. As we shall see in section
  \ref{sec:nonisot:sys}, however, systematic effects in the gas density from
  uncertainties in the metal abundance are non-negligible.}


Using Abel's integral equation, equations (\ref{eq:szrad}) and (\ref{eq:xrad})
can be inverted to obtain joint radial density and temperature profiles:
\begin{equation}
T_e(r)\ n_e(r) = \frac{1}{\pi A_{\mathrm{SZ}}}\ 
\int_{\infty}^r\ \dfrac{\mathrm{d}\Delta T(R)}{\mathrm{d}R} \ 
\dfrac{\mathrm{d}R}{\sqrt{R^2-r^2}};
\label{eq:szabel}
\end{equation} 
\begin{equation}
n^2_e(r) \, \Lambda(T_e(r))  = 4(1+z)^4 
\int_{\infty}^r\ \dfrac{\mathrm{d}S_{\mathrm{X}}(R)}{\mathrm{d}R} \ 
\dfrac{\mathrm{d}R}{\sqrt{R^2-r^2}}.
\label{eq:xabel}
\end{equation} 
Following Yoshikawa \& Suto (1999), we integrate equations (\ref{eq:szabel})
and (\ref{eq:xabel}) numerically by summing in radial bins from
$i_{\mathrm{min}}$ to $i_{\mathrm{max}}$, where $i_{\mathrm{max}}$ is the
index for the outermost bin, and $i_{\mathrm{min}}$ corresponds to $r/D_A$.
We propagate the errors on the density and temperature profiles by a
Monte-Carlo method, the details of which will be discussed in a future
publication (Basu et al. 2009, in preparation). 


\subsubsection{Profile extraction}

The SZE profile is derived as a radial average of the data and errors are
estimated using the same method as in H09.  In order to combine the two data
sets, the X-ray raw mosaic has been smoothed to the APEX-SZ resolution of
1$^{\prime}$.  In doing so, we have neglected the effects of the XMM point
spread function (PSF) and its variation over the field since they are
negligible compared to the APEX-SZ beam.  The X-ray source mask is used to
remove the sub-cluster 8 arcminutes north of Abell 2163, prominent in Figure
\ref{fig:xmm.map}, and the numerous AGN.  A count profile is then extracted
from the background subtracted mosaic and corrected for the local exposure
time and unmasked area, thus converting to surface brightness.  The Poisson
error bars on the counts profiles are re-scaled accordingly. Since the
background and source counts in each annulus are large, the noise can be
considered as Gaussian which justifies the direct background subtraction.
Because of the slight offset between the peaks of the X-ray surface brightness
and the SZE temperature decrement, we have taken the centroid position of the
X-ray map to be the common center in the X-ray/SZE joint analysis because of
the much better signal-to-noise ratio of the X-ray map. 

\subsubsection{Mass estimation method}

Using the SZE and X-ray measurements, we can determine the gas mass and total
mass enclosed within a certain radius assuming hydrostatic equilibrium. This
can be used to compute the gas-to-mass ratio as a function of radial distance,
or can be combined with the weak-lensing data for a more direct determination of
the gas mass fraction. We show the comparison between the gas mass 
fraction values obtained using our non-parametric de-projection method and the 
standard isothermal modeling inside the $r_{500}$ of the cluster.

Computing the gas mass and total mass profiles from isothermal $\beta$ models
is straightforward (e.g. LaRoque et al. 2006). The gas mass is obtained
directly from the electron density profile as
\begin{equation}
M_{\mathrm{gas}}(<r) = 4\pi \ \mu_e n_e(0) m_p D_A^3 
\int_0^{r/D_A} \left(1+ \frac{\theta^2}{\theta_c^2}\right)^{-3\beta /2} \ 
\theta^2 \mathrm{d}\theta,
\label{eq:betamass}
\end{equation} 
where $\mu_e$ is the mean molecular weight per electron, which we assume to be
equal to 1.17 for cosmic abundance of H and He.  For the non-isothermal
analysis, we use the de-projected electron density profile to compute the gas
mass directly as $\rho_{\mathrm{gas}}(r) = \mu_e m_p n_e(r)$.

The total mass, $M_{\mathrm{total}}$, is obtained by solving the hydrostatic
equilibrium (HSE) equation, assuming spherical symmetry, as follows:
\begin{equation}
M_{\mathrm{total}}(<r) = - \frac{k_{\mathrm{B}}T_e(r) \ r}{G \mu m_p} \ 
\left[\frac{d \ln n_e(r)}{d \ln r} + \frac{d \ln T_e(r)}{d \ln r}\right].
\label{eq:totmass}
\end{equation} 
For isothermal modeling, we use the simple analytic form obtained from the
above equation (e.g. Grego et al. 2001)
\begin{equation}
M_{\mathrm{total}}(<r) = \frac{3\beta \ k_{\mathrm{B}}T_e}{G \mu m_p} 
\frac{r^3}{r^2 + r_c^2},
\label{eq:isomass}
\end{equation} 
whereas for non-isothermal modeling we solve equation \ref{eq:totmass}
directly.  The gas mass fraction is in both cases computed as
$f_{\mathrm{gas}}(<r) = M_{\mathrm{gas}}(<r)/M_{\mathrm{total}}(<r)$.

\subsection{Systematics}
\label{sec:nonisot:sys}

{ The dominating source of error in the non-parametric de-projection
  analysis comes from uncertainties in the SZE map at 150 GHz. Intrinsic noise
  properties in the raw map cause the uncertainties in the radial bins to be
  strongly correlated due to noise structures much more extended than the bin width.
  In addition, the deconvolution process discussed in section
  \ref{sec:red:apexsz} not only recovers source signal, but can also
  amplify noise structures in the raw map, thus increasing the uncertainties
  in the radial profile (see Fig.  \ref{fig:aszca:profile}). This
  amplification of noise effectively limits the range of radii for which we
  can make meaningful estimates of the ICM temperature and density.}

{As already discussed in section \ref{sec:iso:sys:pipe}, some modes
  corresponding to large angular scales are irrecoverable by the transfer
  function $-$ in particular, no structures more extended than the path traced by a
  single bolometer during a scan can be recovered. Although the SZE profile is
  well constrained within $r_{500}$, this inherent filtering can lead to
  systematic shifts in temperature and density at larger radii, where the SZE
  signal is potentially underestimated.}

Cavaliere, Lapi and Rephaeli (2005) showed that the Abel deprojection
  method in itself can lead to biases in the estimated temperature and density
  profiles.  However, they also found that this bias is significantly smaller
  than the statistical uncertainty, which they optimistically assumed to be
  1\% on the central value.

{Other systematic uncertainties come from primary CMB signals, modeling of the
  X-ray background, and the metal abundance used in computing the X-ray
  cooling function. We consider each source of systematic uncertainty in what
  follows.  The results are summarized in table
  \ref{table:depsys}. Representative radial scales are taken from the best-fit
  NFW model from weak-lensing analysis of this cluster (Radovich et al. 2008):
  $r_{2500}=3.1^{\prime}$ and $r_{500}=7.6^{\prime}$ .}

\subsubsection{The filter function}

{To address the issue of how irrecoverable signals on large
  angular scales in the SZE map affect the temperature and density profiles,}
we once again carry out a series of simulations (cf.  section
\ref{sec:iso:sys:pipe}), passing artificial models through the reduction
pipeline.  While in section \ref{sec:iso:sys} we considered only how the
signal was affected by the pipeline in terms of fitting a parametric function
out to a certain radius using the transfer function, we now also have to
consider systematic effects from using the transfer function to deconvolve the
map rather than to convolve a model (the latter is much more
straightforward). Lacking a more complete model of the cluster emission, we
again use spherical $\beta$ models, and vary the best-fit parameters from
section \ref{sec:iso} within their 1$\sigma$ errors to test for stability.
After passing the models through the pipeline, we deconvolve them using the
transfer function and compare the resulting profiles with the input models.
While the systematic errors estimated from this method are comparable to the
random errors on the SZE profile inside $r_{500}$, the signal is
systematically lowered by as much as 40\% of the input signal at $r_{200}$.
{Note, however, that at \textit{any} radius considered in our analysis,
  this systematic effect is small compared to the intrinsic statistical error
  originating in the raw map.}

It should be stressed that these error estimates are only meaningful in the
context of the isothermal model.  They are justified, however, by the results
of section \ref{sec:iso}, which indicate that this model indeed provides a
very reasonable fit to the data out to $r_{500}$. In contrast to what was done
in the isothermal analysis, we do not attempt to correct for this effect since
we have no parametric model. Instead, we treat the systematic as an added
uncertainty on the SZE profile, {and propagate this uncertainty through the
  de-projection analysis to estimate the resulting uncertainties in
  temperature and density.}

\subsubsection{Primary CMB}

The systematic uncertainty from the primary CMB signal is estimated using a
method analogous to that described in section \ref{sec:iso:sys}. {After
  a set of simulated primary-CMB maps are subtracted from time stream data and
  passed through the pipeline reduction, deconvolution is performed on each
  map in the set. The different maps are then passed through the de-projection
  analysis to estimate the impact on the derived quantities. It is found that
  the uncertainties are comparable with those of the filter function.}

\subsubsection{X-ray background and metal abundance}

{Although the statistical errors on the recovered temperature and 
density profiles are clearly dominated by the uncertainty in the SZE signal,
several X-ray systematics have to be considered.}

{As mentioned in section \ref{sec:xreduc}, the background model can 
have a significant impact on the X-ray surface brightness at large radii. 
To quantify this further, we fix all the parameters of the background model 
to their best fit value, and allow for an overall change in normalization 
in the instrumental and cosmic background. Identifying the Cash statistic 
with a $\chi^2$ distribution (valid in the large number limit), we derive 
a 1$\sigma$ uncertainty for this normalization factor of +1.9\%/-1.2\%. 
At $r_{500}$, this translates into a 3\% uncertainty in the temperature 
profile and 1\% for the electron density, much below the statistical errors.}

{Our assumption of a fixed metal abundance is another potential source
  of systematics. Line emission is indeed preeminent in the soft energy band
  and this could result in a significantly different value of the cooling
  function required for our de-projection analysis. We have thus investigated
  the impact of a global change in metallicity in the range 0.2 to
  0.8$Z_\odot$. These values correspond to the most extreme 1$\sigma$ bounds
  measured anywhere in the cluster by Snowden et al. (2008) from X-ray
  spectroscopy, yet they only change our results by up to few per cent, again
  smaller than the statistical uncertainties.  The small systematic
  uncertainties are a direct consequence of the high temperatures measured for
  this cluster; line emission at soft energies is indeed much reduced for
  plasma hotter than $\sim$2~keV in which more energetic transitions are
  favored.}

\begin{table}
\begin{minipage}[t]{\columnwidth}
\caption{{Systematic effects in the de-projected temperature and density
  profiles. Statistical errors are indicated for comparison.}}             
\label{table:depsys}      
\centering                          
\renewcommand{\footnoterule}{}  
\begin{tabular}{l c c c c}        
\hline\hline                 
\vphantom{\rule{0pt}{3.5mm}}Source of & \multicolumn{2}{c}{Effect on $T_e$} & \multicolumn{2}{c}{Effect
  on $n_e$} \\
uncertainty & $r_{2500}$ & $r_{500}$ & $r_{2500}$  & $r_{500}$ \vspace*{1mm}\\
\hline
\vphantom{\rule{0pt}{4mm}}Primary CMB & $\pm3.7\%$ & $\pm8.7\%$ & $\pm1.1\%$& $\pm3.3\%$ \\
\vphantom{\rule{0pt}{4mm}}Filter function (SZE map) & $+2.1\%$ & $+8.9\%$ & $+0.3\%$ & $+1.4\%$ \\
\vphantom{\rule{0pt}{4mm}}X-ray background & $^{+0.5\%}_{-0.3\%}$ & $^{+2.4\%}_{-3.1\%}$ &
$^{+0.8\%}_{-0.2\%}$ & $^{+1.1\%}_{-0.3\%}$ \\
\vphantom{\rule{0pt}{4mm}}Metal abundance & $^{+2.2\%}_{-1.3\%}$ & $^{+1.5\%}_{-0.2\%}$ &
  $^{+1.9\%}_{-2.8\%}$ & $^{+2.4\%}_{-3.5\%}$ \vspace*{1.2mm}\\
\hline
\vphantom{\rule{0pt}{4mm}}Total systematic\footnote{Added in quadrature} & $^{+4.8\%}_{-3.9\%}$ & $^{+12.8\%}_{-9.2\%}$ &
$^{+2.4\%}_{-3.0\%}$ & $^{+4.5\%}_{-4.8\%}$ \vspace*{1.2mm} \\
\hline
\vphantom{\rule{0pt}{4mm}}Statistical & $\pm 19.3\%$ & $\pm 43.5\%$ & $\pm 1.1\%$& $\pm 6.0\%$ \vspace*{1mm}\\                                   
\hline
\end{tabular}
\end{minipage}
\end{table}

\subsection{Results of the non-isothermal modeling}

\subsubsection{Density and temperature profiles}
\begin{figure}[t]
  \centering
\includegraphics[width=8cm]{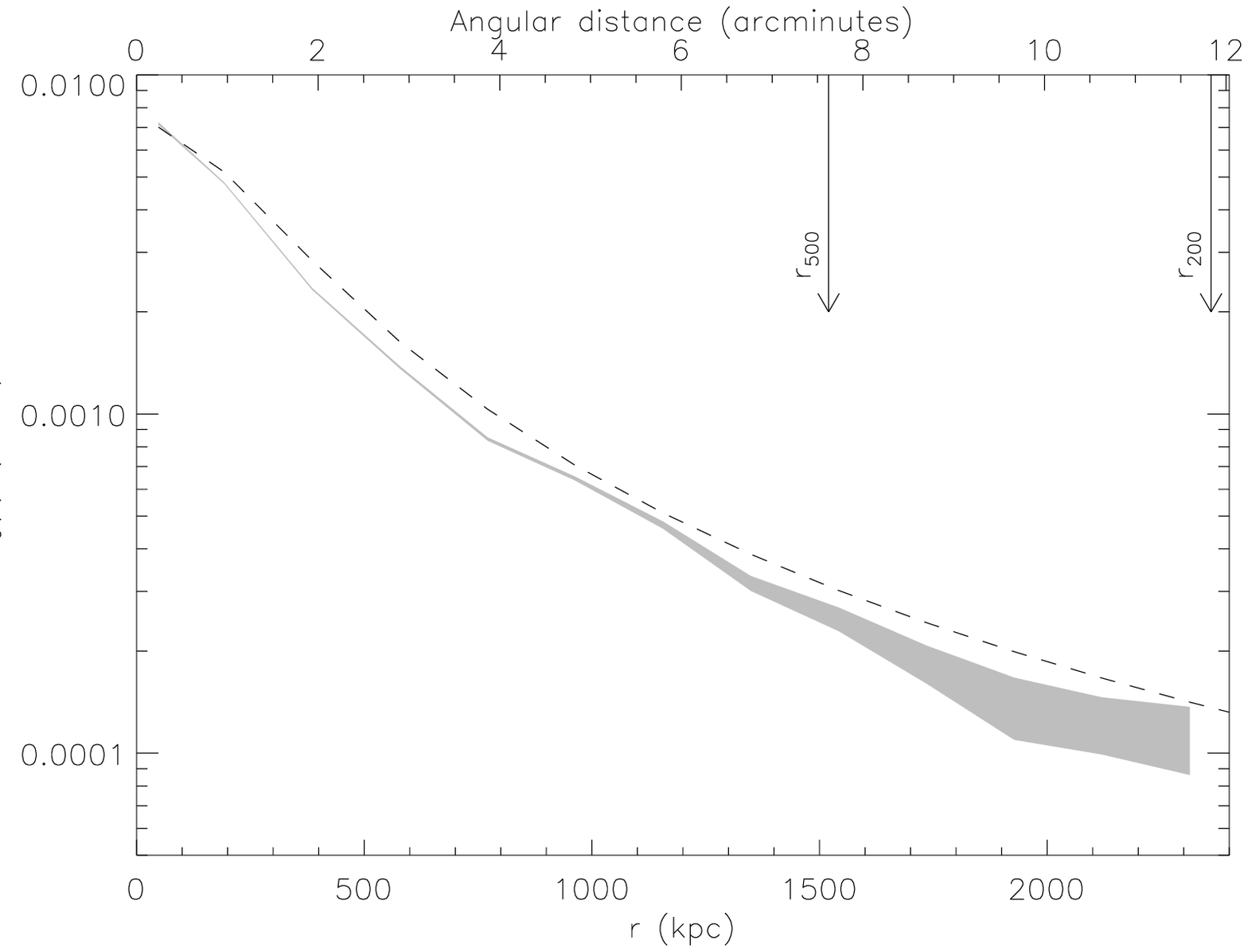} \\
\includegraphics[width=8cm]{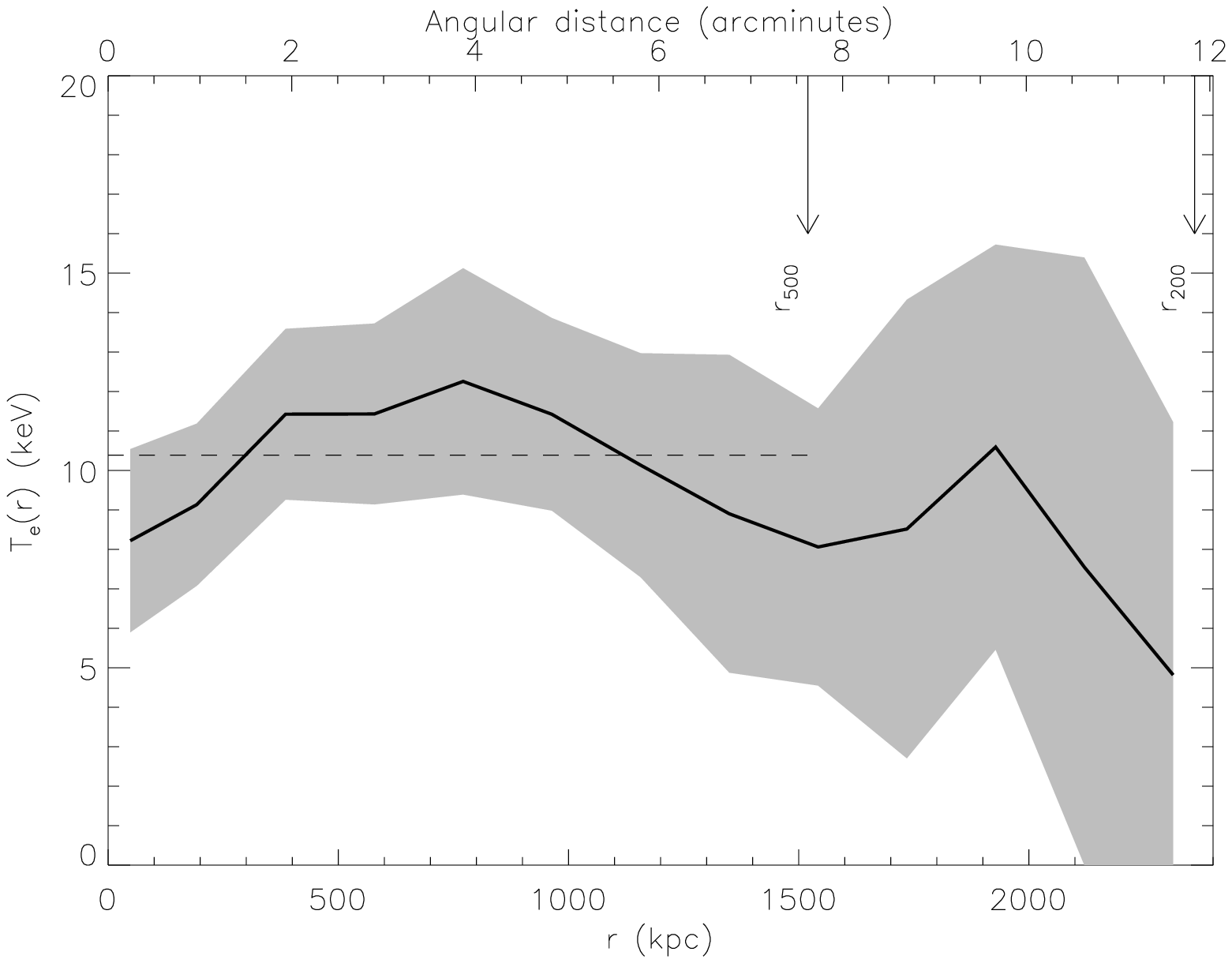}
\caption{{De-projected radial density and temperature profiles of Abell 2163
    from X-ray (XMM-Newton) and APEX-SZ data. } \textit{Top:} {De-projected
    radial density profile (1$\sigma$ confidence region)}. The dashed line is
  the X-ray derived spherical $\beta$ profile ($\beta=0.64$ and
  $\theta_c=91^{\prime\prime}$), normalized to the derived central electron
  density value. \textit{Bottom:} {De-projected temperature profile with
    1$\sigma$ uncertainties}.  The horizontal dashed line shows the best-fit
  isothermal value of
  {$10.4\pm1.4$} keV within $r_{500}$
  ($\sim$7.6$^{\prime}$). Vertical arrows in both plots mark the estimated
  values of $r_{200}$ and $r_{500}$.  
}
\label{fig:numint}
\end{figure}

The radial profiles of the ICM density and temperature, obtained from
equations \ref{eq:szabel} and \ref{eq:xabel}, are shown in Figure
\ref{fig:numint}. 
%
%
The density profile shows very little deviation from an isothermal $\beta$
model inside $r_{500}$ (roughly 1500 kpc), which is expected since the density
values are mostly constrained by the X-ray surface brightness map. In the
[$0.5$-$2$] keV energy band the X-ray surface brightness depends weakly on the
gas temperature; for this reason the de-projected density profile using this
band has a weak dependence on temperature variations. 
%
%
Inside $r_{500}$ the temperature can be fit with a constant (i.e.  isothermal)
value, at 
{$10.4\pm1.4$} keV, marked by the
horizontal dashed line in Fig.\ref{fig:numint}.

{Since no ``de-projected" temperature profile for Abell 2163 from X-ray
  spectroscopic measurements is available in the literature, we compute the
  mean weighted value of the gas temperature along the line of sight to
  compare our results with the published X-ray temperature values.}  
{The projected
  gas temperature is computed} as $T_{\mathrm{proj}} \equiv \int WT dV \ /
\int W dV$, where $T$ is the de-projected gas temperature obtained from Abel
inversion, and $W$ is the weight function. As expected, the effect of
projection is small compared to the errors in our temperature profile. We
compare two different weighing schemes: the standard
emission weight with $W = n^2 \Lambda(T)$,
and the weighing for a
``spectroscopic-like'' temperature as discussed by Mazzotta et al. (2004),
using $W = n^2 T^{-3/4}$.  The difference between these two weighing schemes
is negligible, as can be expected from the slowly varying temperature profile
of A2163 which shows no strong non-isothermal features.

\begin{figure}[ht]
  \centering
\includegraphics[width=8cm]{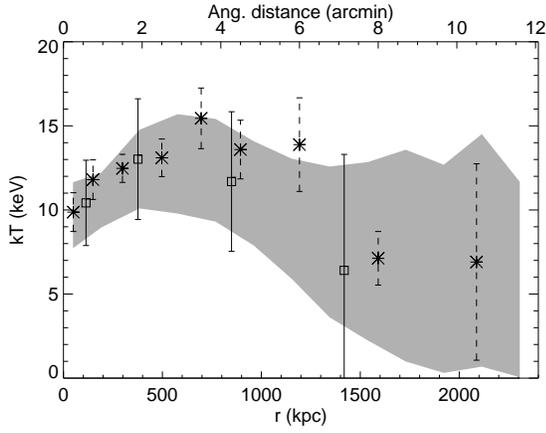} 
\caption{{Comparison of the projected radial temperature profile for
    Abell 2163 with spectroscopic X-ray measurements. The shaded region shows
    the emission-weighted temperature computed from the de-projected
    temperature and density profiles. The squares and crosses are X-ray
    spectroscopic measurements from \textit{Chandra} (Markevitch \& Vikhlinin
    2001) and \textit{XMM-Newton} (Snowden et al. 2008), respectively.}
}
\label{fig:proj}
\end{figure}

{In Figure \ref{fig:proj} we compare the emission-weighted temperature
  with published spectroscopic measurements from \textit{XMM-Newton} and
  \textit{Chandra} observations.}
{The \textit{XMM-Newton} temperature profile indicates a drop in
  temperature in the central region, which may be the remnant of the cold core
  of the original more massive cluster in this merging system. Gas
  temperatures derived from the APEX-SZ measurement are consistent with this
  feature.  In the outskirts, gradually decreasing temperatures with incresing
  radii have been observed for relaxed clusters (e.g. Pratt et al. 2007), and
  is expected from both theory (e.g. Frenk et al. 1999) and numerical
  simulations of cluster models (e.g.  Roncarelli et al.  2006, Hallman et al.
  2007). Although the large uncertainties in the present temperature profile
  beyond $r_{500}$ do not allow any conclusions as to a decrease in
  temperature at large radii, the data are fully consistent with such a
  trend.}

\subsubsection{Gas mass and total mass}
\label{sec:nonisot:gasmass}

\begin{figure}
\centering
\includegraphics[width=8.5cm]{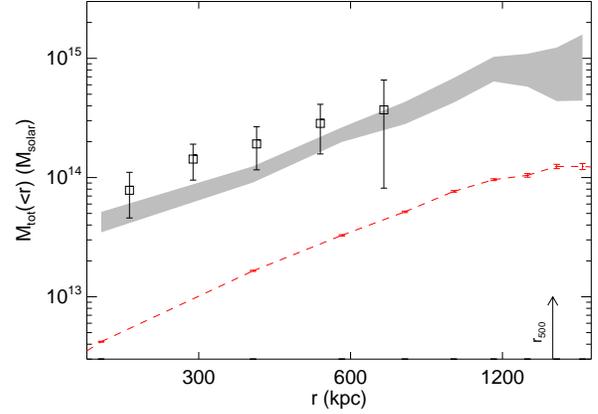}
\includegraphics[width=8.5cm]{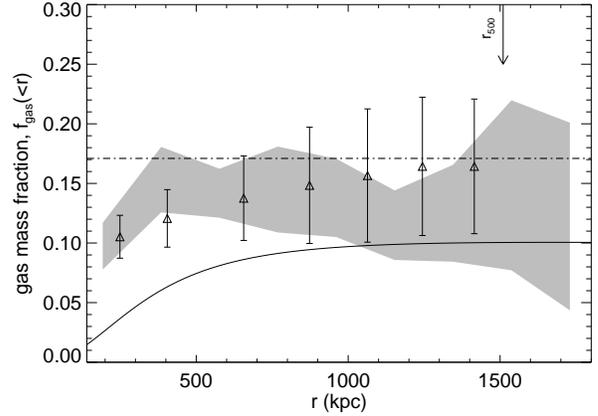}
\caption{\textit{Top:} {Gas mass and total mass in Abell 2163} within $r_{500}$ ($\sim$1500 kpc),
  obtained from the de-projected density and temperature profiles.  
  {The shaded region shows the total mass within $68\%$ confidence
    levels. The dashed line is the gas mass, constrained primarily from the
    X-ray data. The square boxes with error bars are the weak-lensing mass
    profile from Squires et al. (1997).}  \textit{Bottom:} The gas mass
    fraction obtained from the isothermal analysis (solid line) and from the
    non-parametric de-projection (shaded region, $68\%$ CL).  
The triangles with error bars are the results from the
  X-ray analysis of Squires et al.  (1997), and the horizontal dot-dashed line
  represents the cosmic baryon fraction from the WMAP 5-year result (Dunkley
  et al. 2009).}
\label{fig:massfrac}
\end{figure}


The results of the mass analysis are shown in Figure \ref{fig:massfrac}.  We
have used median smoothing to the temperature and density profiles before
computing their derivatives, and limited this analysis to within $r_{500}$
(approximately 1500 kpc) to avoid the large uncertainties at the outer
radii. {The errors in the total mass and gas mass fraction profiles reflect
  the combined random and systematic errors in our analysis.}
We find that the cumulative mass profile obtained under the assumption of HSE
with spherical symmetry is in good agreement with the total mass obtained from
weak lensing (Squires et al. 2007).  The present analysis provides a better
constraint on the total mass profile if the assumption of HSE is
valid.
{The cumulative $f_{\mathrm{gas}}$ values near the cluster center
  obtained from the non-parametric modeling are systematically higher than the
  isothermal $\beta$-model predictions. The drop to zero of the gas mass
  fraction at small radii is a well-known artefact of the isothermal
  $\beta-$modeling of cluster ICM, and our non-parametric profile may be
  indicative of the actual ratio between gas and total gravitational matter
  near the cluster center.  At a radius of 1 Mpc and beyond, the
  $f_{\mathrm{gas}}$ values from both methods agree within 1$\sigma$.}

{The resulting non-parametric gas mass fraction profile for Abell 2163
  is close to the universal baryon fraction obtained from the WMAP 5-year data
  (Dunkley et al. 2009), and is also consistent with the X-ray analysis of
  Squires et al. (1997) for this cluster. The joint SZE/X-ray analysis of the
  gas fraction in clusters using OVRO/BIMA data (LaRoque et al. 2006) also
  yields a value of $f_{\mathrm{gas}}$ in the range $0.15-0.16$ for this
  cluster, using a non-isothermal double $\beta$-model. The latter measurement
  is limited to within $r_{2500}$, and at this radius the SZE-only isothermal
  value is approximately $30\%$ lower than the double $\beta$-model result,
  consistent with the difference found in our analysis.}


It should be noted that the plotted mass and gas mass fractions are cumulative
functions, and hence their values in the outer bins are correlated with the
data in bins closer to the center. Furthermore, the errors in
$f_{\mathrm{gas}}$ are derived from the errors in the total mass and gas mass
profiles, which are not independent. For these reasons, the errors in
$f_{\mathrm{gas}}$ are also correlated.


\section{Constraints from the SZE spectrum}
\label{sec:spectrum}

To compare the SZE decrement/increment values at different frequencies, we
write them in terms of the relative change to the background CMB intensity,
\begin{equation}
\Delta I \ = \ I_0 \ h(x) \ \Omega_{\mathrm{beam}} \ \frac{\Delta T}{T_{\mathrm{CMB}}},
\end{equation} 
where $x\equiv h\nu/kT_{\mathrm{CMB}}$, $I_0\equiv
2(kT_{\mathrm{CMB}})^3/(hc)^2$, $\Omega_{\mathrm{beam}}$ is the beam
equivalent solid angle in steradians, and $h(x)$ relates the frequency
dependence of the SZE expressed in temperature and in intensity as
\begin{equation*}
h(x) = \frac{x^4e^x}{(e^x-1)^2}.
\end{equation*}
The total change of intensity is written as the sum of the thermal and
kinematic components of the SZE as
\begin{equation}
\Delta I \ = \ \Delta I_{\mathrm{T}} \ + \ \Delta I_{\mathrm{K}},
\end{equation} 
where
\begin{equation}
\Delta I_{\mathrm{T}} = I_0 \ y \ h(x) \ f(x, T_e)
\label{eq:thermal}
\end{equation} 
and
\begin{equation}
\Delta I_{\mathrm{K}} = - I_0 \ y \ h(x) \ (m_ec^2/k_{\mathrm{B}}T_e) \ \dfrac{v_{r}}{c},
\label{eq:kinem}
\end{equation} 
with $v_r$ the radial (line-of-sight) peculiar velocity, which is positive for
a receding cluster.

{Relativistic corrections to third order in $k_{\mathrm{B}}T_e/m_ec^2$ are
included in $f(x, T_e)$ according to Sazonov \& Sunyaev (1998) (Note that the
quantity $h(x) \ f(x, T_e)$ is often denoted $g(x, T_e)$ in the literature).}
%
%
Higher order relativistic corrections are negligible given the precision of
the current measurements.  
For the purpose of fitting the SZE spectrum, the Comptonization parameter
{$y$ (defined in equation \ref{eq:compton})} is parameterized in terms
of its central value $y_0$. This allows a simple model where the finite
resolutions of different experiments are accounted for. All the spectral data
are obtained with a radial fit using spherical $\beta$ models, and accounting
for beam dilution. For APEX-SZ and LABOCA, we use the central
decrement/increment values, corrected for dust contamination, derived in
section \ref{sec:iso}.  The data used for the fit to the SZE spectrum are
given in Table \ref{tab:spectrum}.

Due to the degeneracies between velocity, temperature and Comptonization, the
present data is not sufficient for a simultaneous constraint of all three
parameters. Instead, we fix the ICM temperature to 
{10.4}
keV from the joint X-ray and SZE analysis of section \ref{sec:nonisot}, and
perform a least squares fit in $y_0$ and $v_r$. We also fit these parameters
using alternative values of the temperature in the range $8$-$14$ keV.

To estimate the errors in the fitted parameters, we perform a Monte Carlo
simulation, in which 10,000 artificial data sets are created from the actual
spectral data, adding random Gaussian offsets with the amplitude of the
statistical noise at each frequency. To model the effect of the systematics,
we also add random offsets from the most important systematic components from
section \ref{sec:iso:sys}, with the proper correlations between the different
frequencies taken into account.  Dust and primary CMB signals are scaled to
the estimated levels derived in section \ref{sec:iso} at 150 and 345 GHz and
to the levels derived by LaRoque et al.  (2002) at the other frequencies. To
account for primary calibration uncertainties, we assume a 5\% calibration
uncertainty across the spectrum.

\begin{figure}
  \centering \includegraphics[width=8cm]{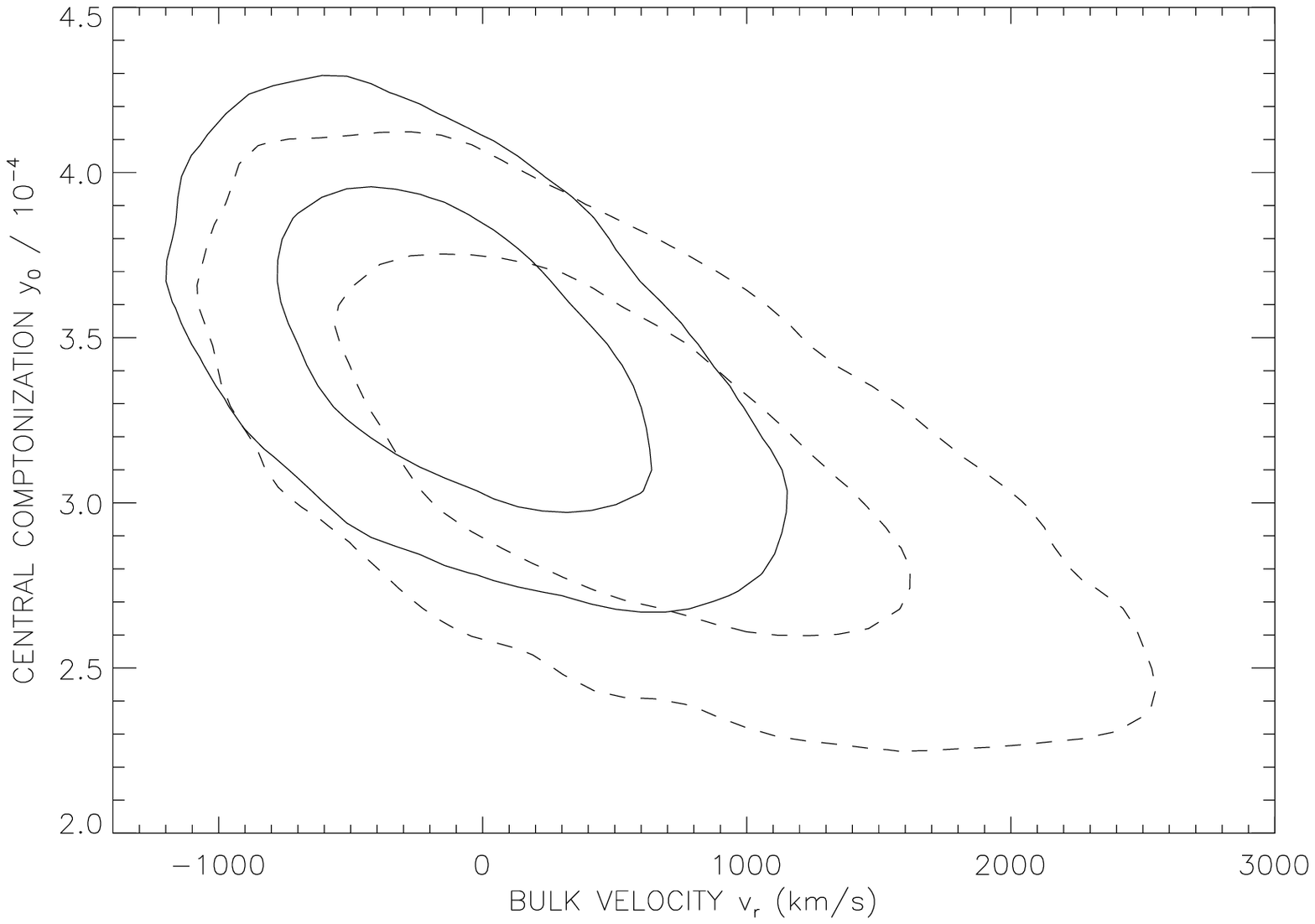}
  \centering \includegraphics[width=8cm]{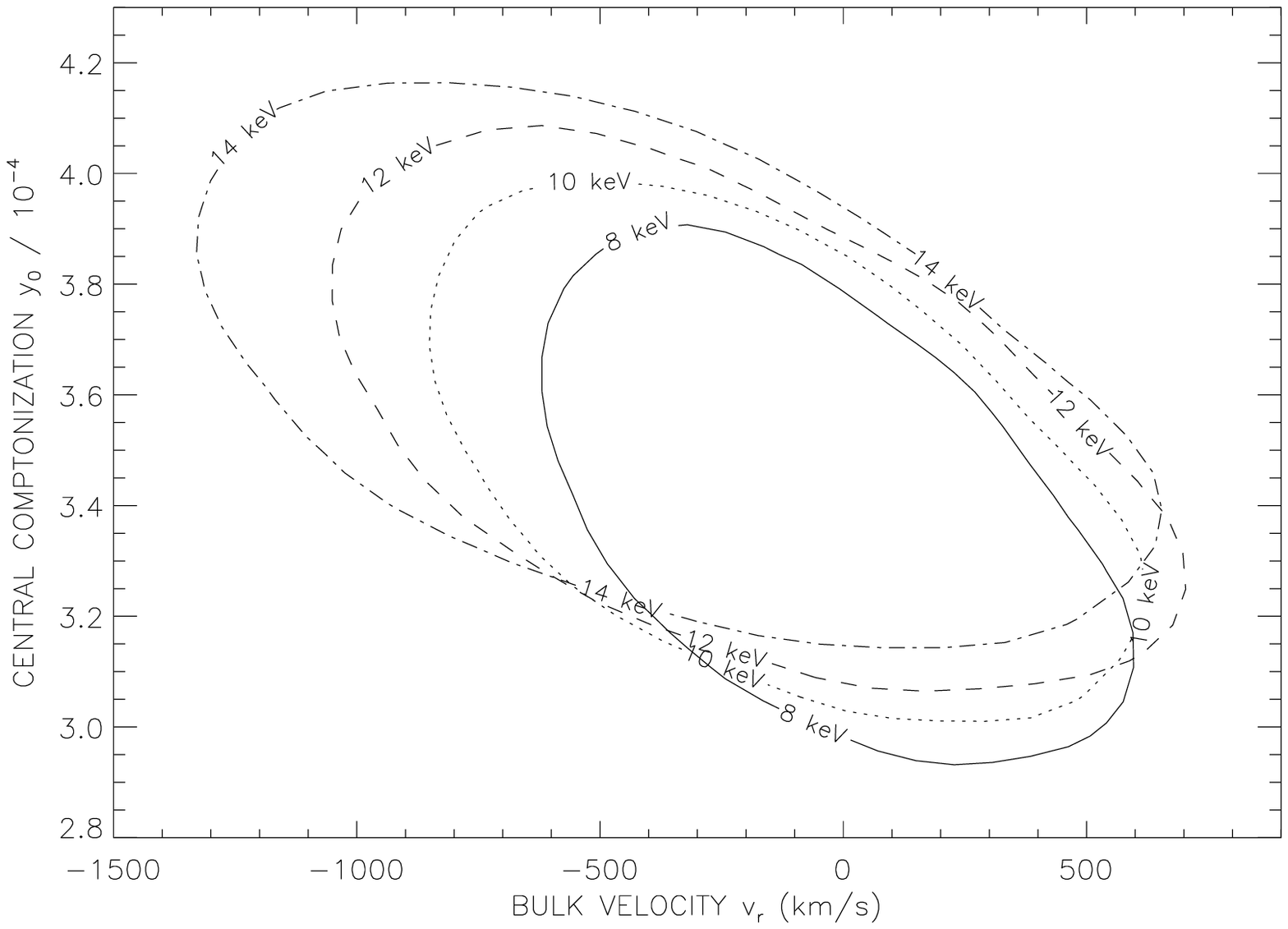}
  \caption{{Constraints in the $v_r-y_0$ parameter space} of Abell
    2163 with priors on the ICM temperature, including systematic
    uncertainties. \textit{Top:} solid contours indicate the 68\% and 95\%
    confidence regions using all available data and assuming $kT_e = $
    {10.4} 
    keV.
    The dashed contours correspond to excluding the present (APEX-SZ and
    LABOCA) measurements.
    \textit{Bottom:} 68\% confidence regions for a range of ICM temperatures
    (indicated) using all available data.}
\label{fig:bananas}
\end{figure}

\begin{figure}
  \centering \includegraphics[width=8cm]{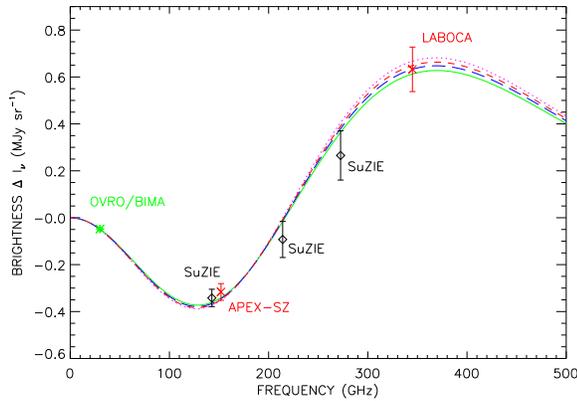}
\caption{{SZE spectrum of Abell 2163} (points) and best-fit models using different
  priors on the ICM temperature: 8 keV (solid line), 10 keV (long-dashed
  line), 12 keV (short-dashed line) and 14 keV (dotted line).}
\label{fig:spectrum}
\end{figure}

The resulting error estimates, including systematics, are shown in Figure
\ref{fig:bananas}. The fitted radial velocity of the cluster is consistent
with zero, in agreement with and marginally improving the constraints of
LaRoque et al. Specifically, we find 
{$v_r=-140\pm460$} km/s and 
{$y_0=3.42\pm0.32 \times 10^{-4}$}, excluding systematic
effects. The fit is dominated by random noise; we find that including the
systematic errors in the modeling increases the error in $y_0$ by 12\% and in
$v_r$ by 18\%. The SZE spectra resulting from the various fits are indicated
in Figure \ref{fig:spectrum}.

\begin{table}
  \begin{center}
    \setlength{\belowcaptionskip}{10pt}
    \caption{SZE decrement/increment measurements used for the spectral
      fit.} 
    \label{tab:spectrum}    
    \begin{tabular}{l l r@{}l}
      \hline\hline
      Wavelength (mm) & Instrument & $\Delta$ & $I \,$  (MJy sr$^{-1}$) \\
      \hline
      10 & OVRO/BIMA$^{1}$ & $-$0&.043$\pm 0.005 $ $^{4}$  \\
      2.1 & SuZIE$^{2}$ & $-$0&.342$\pm 0.033$ $^{3,4}$ \\
      2.0 & APEX-SZ & $-$0&{.317}$\pm 0.035$ \\
      1.4 & SuZIE$^{2}$ & $-$0&.093$\pm 0.069$ $^{3,4}$ \\
      1.1 & SuZIE$^{2}$ & 0&.266$\pm 0.095$ $^{3,4}$ \\
      0.86 & LABOCA & 0&.633$\pm 0.094$  \\
      \hline
    \end{tabular}
    \\
  \end{center}
$^1$ LaRoque et al. (2002) \\
$^2$ Holzapfel et al. (1997) \\
$^3$ dust-corrected intensities from LaRoque et al. (2002) \\
$^4$ calibration corrected according to Hill et al. (2009)
\end{table}

\section{Conclusions}
\label{sec:concl}


\begin{enumerate}
  
\item We present SZE maps of the galaxy cluster Abell 2163 at two frequencies,
  showing the SZE decrement at 150 GHz from observations with the APEX-SZ
  bolometer camera, and the SZE increment at 345 GHz from observations with
  the LABOCA bolometer camera. The 345 GHz measurement is the highest
  resolution SZE image for this cluster to date, and the first large-area ($>
  10^{\prime}$) imaging of a galaxy cluster at sub-mm wavelengths.
  
\item An isothermal modeling of the SZE yields results consistent with the
  X-ray derived isothermal model fits, implying that the large-scale
  properties of the cluster, under the assumption of spherical symmetry, are
  well represented by such models.
 
\item Using the APEX-SZ map in conjunction with XMM-Newton X-ray data we
  derive the (de-projected) radial density and temperature profile of the ICM
  under the assumption of spherical symmetry. 
The projected gas temperature profile is {found to be
    fully consistent with previous X-ray spectroscopic measurements of
    Markevitch \& Vikhlinin (2001) using \textit{Chandra} data, and Snowden et
    al. (2008) using \textit{XMM-Newton} data.}  
   
\item The total mass profile of the cluster is obtained under the assumption
  of hydrostatic equilibrium. The resulting profile agrees well with the
  weak-lensing mass profiles from previous works (Squires et al. 1997,
  Radovich et al. 2008), and extends the profile out to $r_{500}$.
  
\item The 
{gas mass fraction of the ICM} obtained from our 
non-parametric analysis are consistent with the previous X-ray mesurements
(Squires et al. 1997) and SZE/X-ray joint analysis with double $\beta$-model
(LaRoque et al. 2006).  
We do not see any
significant trend for increasing ICM baryonic fraction from $r_{2500}$ to
$r_{500}$ of the cluster.
  
\item Using isothermal fits to the LABOCA and APEX-SZ measurements, we
  constrain the line-of-sight peculiar velocity of the cluster and the central
  optical depth to inverse Compton scattering, using the temperature obtained
  from the de-projection analysis. We find a peculiar velocity
  {$v_r=-140\pm460$} km/s, {consistent with zero},
  and a central Comptonization 
  {$y_0=3.42\pm0.32 \times 10^{-4}$}.

\end{enumerate}

\begin{acknowledgements}
  
  We thank the APEX staff for their unfailing support during the APEX-SZ and
  LABOCA observations. We thank Wilhelm Altenhoff for fruitful discussions on
  the calibration of the millimeter data, and Fred Schuller for helping with
  the LABOCA data taking.
  {This work has been partially supported by the DFG Priority Program
    1177.}  MN acknowledges support for this research through a stipend from
  the International Max Planck Research School (IMPRS) for Radio and Infrared
  Astronomy at the Universities of Bonn and Cologne. FP acknowledges support
  from the DfG Transregio Programme TR33. {NWH acknowedges support from
    an Alfred P. Sloan Research Fellowship.}
  APEX is a collaboration between the Max-Planck-Institut
  f\"ur Radioastronomie, the European Southern Observatory, and the Onsala
  Space Observatory. APEX-SZ is funded by the National
  Science Foundation under Grant No.  AST-0138348.
  XMM-Newton is an ESA science mission with instruments and contributions
  directly funded by ESA Member States and the USA (NASA). The XMM-Newton
  project is supported in Germany by the Bundesministerium f\"ur Wirtschaft
  und Technologie/Deutsches Zentrum f\"ur Luft- und Raumfahrt (BMWI/DLR, FKZ
  50 OX 0001), the Max-Planck Society and the Heidenhain-Stiftung.

\end{acknowledgements}

\end{document}